\documentclass[twocolumn]{aastex631}
\usepackage{graphicx} 
\usepackage{booktabs}
\usepackage{array}

\usepackage{etoolbox}
\BeforeBeginEnvironment{document}{%
  \expandafter\let\csname longtable*\endcsname\relax
  \expandafter\let\csname endlongtable*\endcsname\relax
}
\usepackage{subcaption}

\usepackage{natbib}
\usepackage{float}

\usepackage{lipsum}

\usepackage{xcolor}
\usepackage{amsmath}

 \usepackage{hyperref}

\usepackage{multirow}
\usepackage{orcidlink} 
\newcommand{\gammasource}{$\gamma$}
\usepackage{textcomp}
\usepackage{gensymb}  

\newcommand{\chandra}{\textit{Chandra}}

\usepackage{placeins}

\submitjournal{ApJ}

\accepted{August 22, 2025}

\begin{document}


\title{Detection of Diffuse Radio Emission inside the Supernova Remnant G338.3$-$0.0 associated with the \gammasource-ray Source HESS J1640$-$465}

\author[0000-0002-4441-7081]{Moaz Abdelmaguid} 
\affil{Department of Physics, New York University, 726 Broadway, New York, NY 10003, USA}
\affil{New York University Abu Dhabi, PO Box 129188, Abu Dhabi, UAE}
\affil{Center for Astrophysics \& Space Science (CASS), NYU Abu Dhabi, PO Box 129188, Abu Dhabi, UAE}

\author[0000-0003-4679-1058]{Joseph D. Gelfand}
\affil{New York University Abu Dhabi, PO Box 129188, Abu Dhabi, UAE}
\affil{Center for Astrophysics \& Space Science (CASS), NYU Abu Dhabi, PO Box 129188, Abu Dhabi, UAE}
\affil{Center for Cosmology and Particle Physics (CCPP, Affiliate), New York University, 726 Broadway, New York, NY 10003, USA}

\correspondingauthor{Moaz Abdelmaguid}
\email{m.abdelmaguid@nyu.edu}

\author[0000-0002-2312-8539]{J. A. J. Alford}
\affil{New York University Abu Dhabi, PO Box 129188, Abu Dhabi, UAE}
\affil{Center for Astrophysics \& Space Science (CASS), NYU Abu Dhabi, PO Box 129188, Abu Dhabi, UAE}

\begin{abstract}

We report the discovery of diffuse radio emission within SNR G338.3$-$0.0 using new MeerKAT observations at 816\,MHz and 1.4\,GHz.~The radio emission spatially overlaps with the X-ray pulsar wind nebula (PWN) powered by PSR~J1640$-$4631 and the GeV/TeV $\gamma$-ray source HESS\,J1640$-$465. The morphology of this radio emission is centrally peaked and its extent is well-contained within the SNR shell.~A lack of mid- and far-infrared counterparts and the absence of catalogued H\,\textsc{ii} regions argues against a thermal origin,  while the morphology and radial profile are suggestive of a PWN origin powered by PSR~J1640$-$4631.~Under this assumption, we use a one-zone, time dependant model to reproduce the size and broadband (radio, X-ray, and $\gamma$-rays) spectral energy distribution of the PWN. The modelling and broadband properties of this PWN suggests it is currently interacting with the reverse shock within its host SNR. This evolutionary  stage is associated with particles escaping the PWN and entering the ISM, suggesting this object may be an important source of Galactic PeV e{$^\pm$}.

\end{abstract}


\section{Introduction} \label{sec:intro}

 $\gamma$-ray observations strongly suggest that supernova remnants (SNRs) and pulsar wind nebulae (PWNe) are two of the primary particle accelerator sources in our galaxy \citep{2022hxga.book...52V, 2006ARA&A..44...17G}.~In PWNe, the $\gamma$-ray emission is predominantly leptonic, driven by relativistic electrons scattering low-energy photons to TeV energies via inverse Compton scattering (ICS) \citep{1996A&AS..120C.453A, 2006ARA&A..44...17G}.~In SNRs, both hadronic and leptonic processes can produce $\gamma$-rays:~hadronic emission arises from shock-accelerated protons interacting with surrounding material, producing neutral pions that decay into $\gamma$-rays, while leptonic emission involves electrons upscattering ambient photon fields \citep{2022hxga.book...52V}.~A notable example, HESS J1640$-$465, illustrates the complexity of disentangling these origins, with TeV emission potentially arising from the leptonic PWN or the SNR shell via hadronic and/or leptonic mechanisms. 


HESS~J1640$-$465 is a powerful source of very high-energy (VHE; \( E_{\gamma} > 0.1 \)~TeV) $\gamma$-rays, first identified during the High Energy Stereoscopic System (H.E.S.S.) Galactic Plane Survey (HGPS) with a spatial extent of $2\farcm7$ \citep{2006ApJ...636..777A}. Subsequent H.E.S.S. observations revealed a more extended morphology, with a gaussian width of  $\sigma = 4\farcm32 \pm 0\farcm18$ \citep{2014MNRAS.439.2828A}.~Its TeV emission is spatially coincident with G338.3$-$0.0; a shell-like radio SNR with a $4^\prime$ radius \citep{1996A&AS..118..329W}. \textit{XMM-Newton} observations revealed an asymmetric and slightly extended X-ray emission near the SNR's center, suggesting the presence of a PWN \citep{2007ApJ...662..517F}.~Subsequent \textit{Chandra} observations confirmed the existence of an extended nebula surrounding a point-like X-ray source, a potentially associated pulsar \citep{2009ApJ...706.1269L}.~No X-ray emission from the SNR shell has been detected, likely due to the low temperature of the ejecta and/or the high Hydrogen column density along the line of sight \citep{2009ApJ...706.1269L}.~Furthermore, \cite{2009ApJ...706.1269L} estimated a distance of 8$-$13~kpc to this system based on 21 cm H\,\textsc{i} absorption, suggesting it is a very luminous galactic VHE $\gamma$-ray source, with $L(0.2-10\ {\rm TeV})=2.8\times10^{35}\,(d/12\ {\rm kpc})^2$~erg~s$^{-1}$.~Subsequent observations with the Nuclear Spectroscopic Telescope Array (\textit{NuSTAR}; \citealt{2013ApJ...770..103H}) by \citet{2014ApJ...788..155G} revealed X-ray pulsations from a point source within HESS~J1640$-$465, leading to its identification as a young, energetic pulsar: PSR~J1640$-$4631. The pulsar has a spin period of 206 ms, a spin-down luminosity of \( \dot{E} = 4.4 \times 10^{36} \) erg s\(^{-1} \), and a characteristic age of $\tau_c \approx 3.4\;$kyr.~\cite{2016ApJ...819L..16A} measured its braking index to be p = $3.15 \pm{0.03}$, marking it as the first pulsar with a braking index $>3$.

Radio observations of SNR G338.3$–$0.0 using the Giant-Metrewave Radio Telescope (GMRT) at 235, 610 and 1280 MHz and the Australia Telescope Compact Array (ATCA) at 1290 and 2300 MHz did not detect a compact radio counterpart at the position of the X-ray PWN, yielding an upper limit on the 610 MHz flux density of the putative pulsar of $S_{610} \sim6~\mathrm{mJy}$ \citep{2011_castellati}.

In terms of $\gamma$-ray observations, analysis of data from the Fermi Large Area Telescope (LAT; \citealt{fermi_lat}) reveals a GeV $\gamma$-ray source, 1FGL~J1640.8$-$465, with a relatively soft spectral index of $\Gamma = 2.3 \pm 0.1$, which is spatially coincident with HESS~J1640$-$465 \citep{2010ApJ...720..266S}.~The subsequent discovery of a nearby TeV source, HESS~J1641$-$463 \citep{2014ApJ...794L...1A}, revised the 0.1$-$300~GeV spectral index of HESS~J1640$-$465 to a harder value of $\Gamma = 1.99 \pm 0.04$ \citep{2014ApJ...794L..16L}.~A follow-up analysis by \citet{2018ApJ...867...55X} in the 10$-$500~GeV range revealed an even harder spectrum, with $\Gamma = 1.42 \pm 0.19$.~More recently, \citet{2021ApJ...912..158M} analyzed eight years of \textit{Fermi}-LAT data using the updated 4FGL catalog \citep{2020ApJS..247...33A}, identifying 4FGL~J1640.6$-$4632 as spatially coincident with HESS~J1640$-$465.~This study incorporated improved diffuse emission models and accounted for contamination from the nearby source 4FGL~J1639.8$-$4642, resulting in a power-law index of $\Gamma = 1.8 \pm 0.1_{\mathrm{stat}} \pm 0.2_{\mathrm{syst}}$  over the 200~MeV to 1~TeV range. 

Several studies have modeled the $\gamma$-ray emission from HESS~J1640$-$465, exploring both leptonic and hadronic scenarios.~\cite{2010ApJ...720..266S} proposed a leptonic model where ICS scattering by electrons in a low magnetic field environment explains the GeV$-$TeV emission.~\cite{2014ApJ...788..155G} then demonstrated that a time-dependent PWN evolution model with a rapidly spinning pulsar could reproduce the broadband emission.~Using updated Fermi-LAT data, \citet{2018ApJ...867...55X} also reproduced the broadband emission within a leptonic PWN framework.~However, alternative hadronic interpretations have also been explored, where the broadband emission from HESS~J1640$-$465 arises from proton-proton interactions in the dense H\textsc{ii} region located near the SNR’s northwestern shell---where the TeV emission overlaps significantly---as initially proposed by \cite{2014MNRAS.439.2828A}, and then by \cite{2015ApJ...812...32T}, \cite{2016A&A...589A..51S}, and \cite{2017MNRAS.464.3757L}. Recently, \citet{2021ApJ...912..158M} examined both leptonic and hadronic scenarios and found that while either could reproduce the observed emission, the hadronic case required an unusually high electron-to-proton ratio (\( K_{\text{ep}} \gtrsim 5 \times 10^{-2} \)), exceeding typical values (\( K_{\text{ep}} \sim 10^{-5} \text{--} 10^{-2} \)) inferred from other SNRs \citep{2013MNRAS.434.2748C, 2011ApJ...734...85C}.

\begin{table*}[htbp]
    \caption{A Summary of the positions \& sizes of HESS J1640$-$465 and nearby sources}
    \centering
    \begin{tabular}{ccccc}
        \toprule
        \toprule
        Region & R.A (J2000) & Dec (J2000) & Size (Arcmin) & Citation \\
        \midrule
        HESS J1640$-$465 (H.E.S.S)& 16\textsuperscript{h} 40\textsuperscript{m} 38.0\textsuperscript{s} & \( -46^\circ\ 34'\ 23.0'' \) & $4^\prime.3 \pm 0.2$ & \cite{2014MNRAS.439.2828A}   \\
        HESS J1640$-$465 (\textit{Fermi}) & 16\textsuperscript{h} 40\textsuperscript{m} 40.8\textsuperscript{s} & \( -46^\circ\ 32'\ 60.0'' \) & $3^\prime.2 \pm 0.7$ &  \cite{2021ApJ...912..158M} \\
        X-ray PWN (\textit{NuSTAR}) & 16\textsuperscript{h} 40\textsuperscript{m} 42.1\textsuperscript{s} & \( -46^\circ\ 31'\ 47.8'' \)  & $1^\prime.7 \times 1^\prime.3$  & \cite{2014ApJ...788..155G} \\
        X-ray Pulsar (PSR J1640$-$4631) & 16\textsuperscript{h} 40\textsuperscript{m} 43.5\textsuperscript{s} & \( -46^\circ\ 31'\ 35.4'' \)  & - & \cite{2009ApJ...706.1269L} \\
        Interior Radio Diffuse Emission & 16\textsuperscript{h} 40\textsuperscript{m} 49.3\textsuperscript{s} & \( -46^\circ\ 32'\ 40.0'' \) & $2^\prime.0 \times 1^\prime.65$ & This Work \\
        Radio PS & 16\textsuperscript{h} 40\textsuperscript{m} 48.0\textsuperscript{s} & \( -46^\circ\ 31'\ 58'' \)  & - &  This work \\
        Radio SNR G338.3$-$0.0 & 16\textsuperscript{h} 40\textsuperscript{m} 54.0\textsuperscript{s} & \( -46^\circ\ 32'\ 40'' \) & $4^\prime.0$ & \cite{2011_castellati} \\
        \bottomrule
    \end{tabular}
\tablecomments{The listed sizes represent radii for all sources except the X-ray PWN and the interior radio diffuse emission, for which semi-major and semi-minor axes are provided. }
\label{tab:regions}%
\end{table*}

In a more recent study, \cite{2023ApJ...946...40A} predicted that, if present, the radio PWN should have flux densities of $S_{800} \sim 100$ mJy and $S_{1300} \sim 80$ mJy. Although no radio counterpart at the position of the X-ray PWN had been detected \citep{2011_castellati}, some extended radio emission was observed within the SNR (see Figure 2 in \citealt{2011_castellati}).~To further investigate this emission, we conducted a follow-up multi-frequency observation of the region surrounding HESS J1640$-$465 with the MeerKAT radio telescope to search for the possible existence of a radio counterpart to the X-ray PWN.

The paper is structured as follows. In \S\ref{sec:observations}, we describe the MeerKAT observations, including the observational setup (\S\ref{sec:2.1}) and data reduction (\S\ref{sec:2.2}). In \S\ref{sec:results}, we present the methods used to analyze the data and summarize the main observational results.~In \S \ref{sec:interior_prop}, we discuss the properties of the newly detected radio diffuse emission, and in \S \ref{sec:nature}, we examine its possible physical origin. This is followed by new broadband spectral modeling of the source, and a discussion of its implications in \S\ref{sec:modelling}.~Finally, in \S \ref{sec:summary}, we summarize our main findings and outline future directions.

\newpage

\section{MeerKAT Radio Observations } \label{sec:observations}

\subsection{Observation Setup} \label{sec:2.1}

We observed the field of view (FoV) around HESS J1640$-$465 with the MeerKAT radio telescope in the UHF band (544$-$1088 MHz) with an angular resolution $\theta_{\rm res}$ $\sim6.\!\!^{\prime \prime}9$ and the L band (856$-$1712 MHz) with an $\theta_{\rm res}$$\sim3.\!\!^{\prime \prime}5$. Each band was divided into 4096 spectral channels with an integration time of 8 s. The MeerKAT array was pointed at the location of the source: RA: 16\textsuperscript{h} 40\textsuperscript{m} 41\textsuperscript{s} \& Dec:  \( -46^\circ\ 32'\ 31'' \) (J2000). The UHF band (544$-$1088 MHz) observation was carried out on 2024 October $11^{th}$ using 59 antennas at a central frequency of $\nu $= 816 MHz. J1932$-$6342 was observed for 10 minutes as the bandpass, flux and delay calibrator, J1744$-$5144 was observed for 1200 seconds as a gain calibrator and the source was osberved for a total of 20 minutes. The L-band (856$-$1712 MHz) observation was carried out on 2024 October  $13^{th}$ using 60 antennas at a central frequency of $\nu $= 1284 MHz. The observational set-up was the same as the UHF band. 

\subsection{Data Reduction} \label{sec:2.2}
 We converted the observations from the MeerKAT Visibility Format (MVF) into \texttt{CASA} Measurement Set (MS) format (\citealt{2007ASPC..376..127M}; \citealt{2022PASP..134k4501C}) using the \texttt{katdal} package, available online in the MeerKAT archive. We also used its "Default Calibrated" option, which applies all the calibration solutions (bandpass, delay and gain) found by the SARAO Science Data Processor (SDP) calibration pipeline, which uses the model of PKS B0407--658 by \citet{Hugo2021} as its flux density scale. For the UHF band, we analyzed the data products from the SDP pipeline directly, while we re-imaged the L-band data using different weighting settings, which will be discussed in section \S \ref{sec:l_band}. We used the \texttt{MIRIAD} software  \citep{1995ASPC...77..433S} to carry out the data analysis and flux calculations.

\begin{figure}
    \centering
    \includegraphics[width=1.05\linewidth]{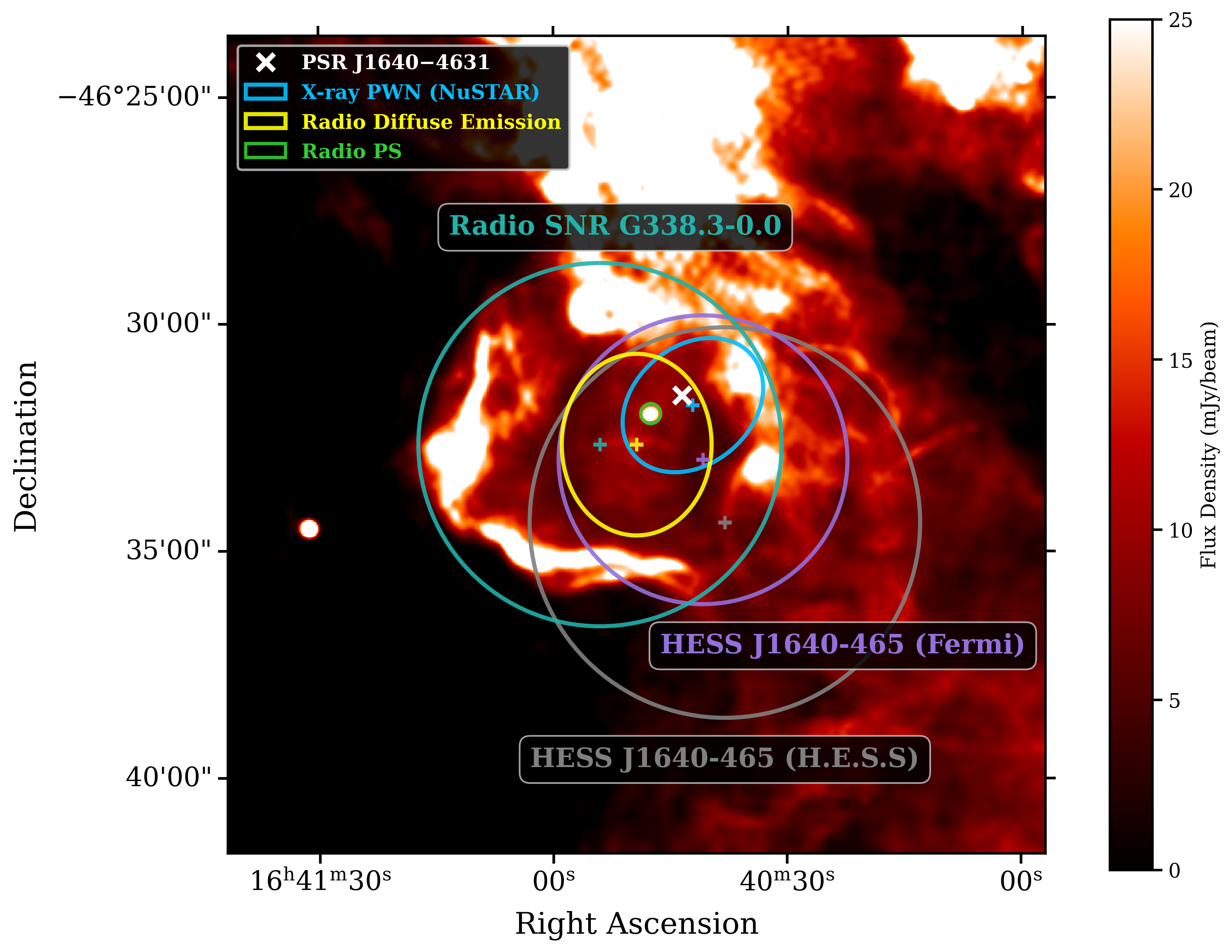}
    \caption{The 816 MHz MeerKAT image of the region around HESS J1640$-$465 overlaid with the extent of the H.E.S.S. source (grey; \citealt{2014MNRAS.439.2828A}), its Fermi-LAT counterpart (purple; \citealt{2021ApJ...912..158M}), along with the radio SNR G338.3$-$0.0 (teal; \citealt{2011_castellati}), a radio point source (PS) within the SNR (green), the X-ray pulsar PSR J1640$-$4631 (white ``x''; \citealt{2009ApJ...706.1269L}) and the interior radio diffuse emission (yellow). The centroid of each source is indicated with a cross of its respective color.~The coordinates and sizes of these objects are listed in Table \ref{tab:regions}. The intensity scale of the colormap is linear and in the units of mJy beam$^{-1}$.}
    \label{fig:new_figure}
\end{figure}

\begin{figure}[htbp]
    \centering
    \includegraphics[width=1.0\linewidth]{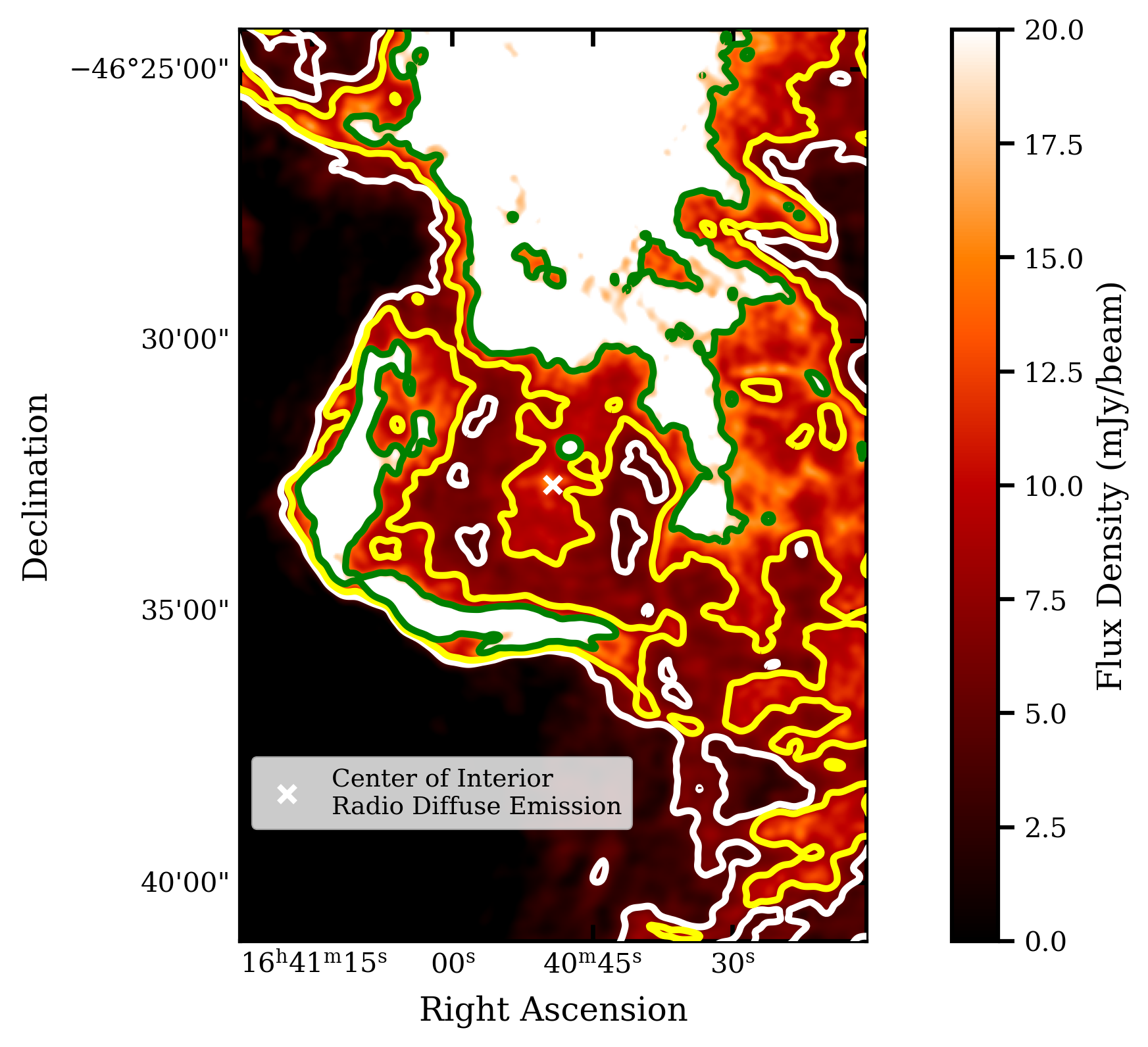}
    \caption{Contour map of the FoV around SNR G338.3$-$0.0 at 816 MHz.~The contour levels (white, yellow and green) correspond to the  3$\sigma$,  5$\sigma$ and 10$\sigma$ levels respectively, where $\sigma$ represents the rms noise in our observation, estimated to be 1.654 mJy beam$^{-1}$. The intensity scale of the colormap is linear and in the units of mJy beam$^{-1}$. The center of interior radio diffuse emission is marked by a white cross and its coordinates are listed in Table \ref{tab:regions}.}
    \label{fig:contours}
\end{figure}

\begin{figure*}[htbp]
        \centering
        \includegraphics[width=1.0\linewidth]{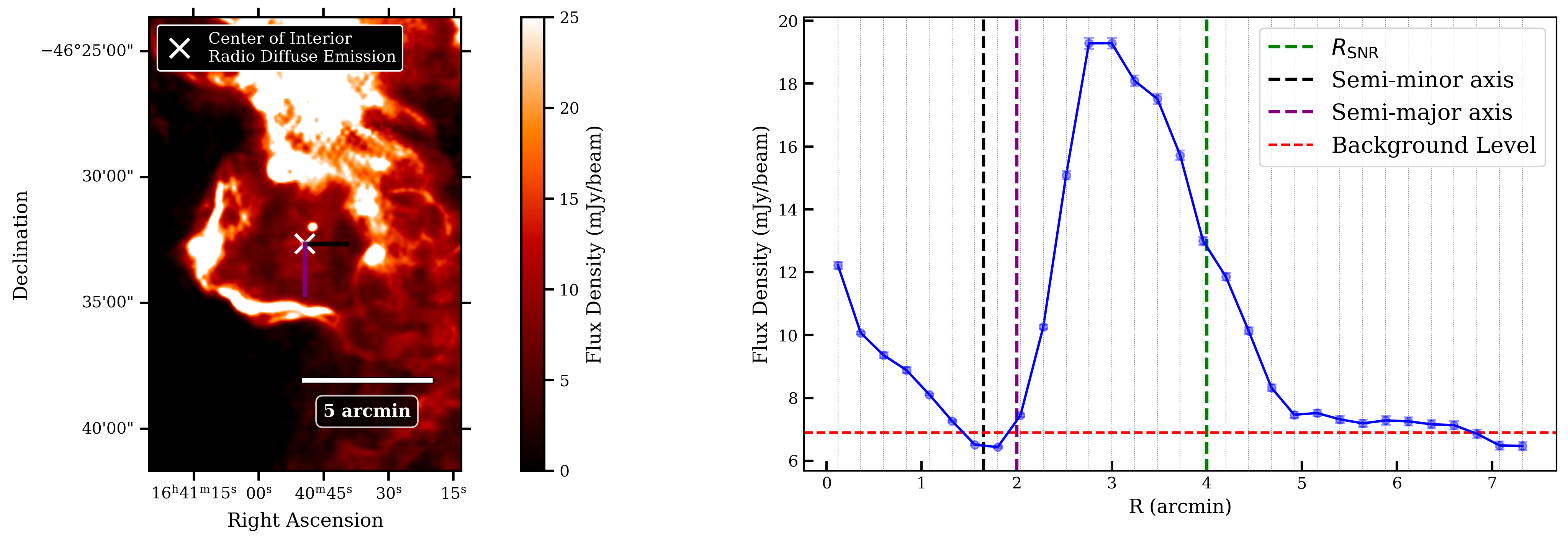}
\caption{Radial profile of the surface brightness toward SNR G338.3$-$0.0. \textit{Left Panel:} Intensity map with the white cross marking the center of the interior radio diffuse emission region.~The black and purple dashed segments correspond to the semi-minor and semi-major axes of the same region, respectively. The intensity scale of the colormap is linear and in the units of  mJy beam$^{-1}$. \textit{Right Panel:} Radial profile of the interior radio diffuse emission region and SNR G338.3$-$0.0 averaged over the entire shell. The black and purple dashed lines represent the semi-minor and semi-major axes of the interior radio diffuse emission region, while the dashed green line represent the estimated radius of SNR G338.3$-$0.0 \citep{2011_castellati}. The red dashed line indicates the estimated background level (see \S\ref{sec:background})}. 
\label{fig:radial_profile}
\end{figure*}

\section{Data Analysis $\&$ Results} \label{sec:results}

\subsection{UHF Band Emission} 
\label{sec:UHF_band_analysis}
Figure~\ref{fig:new_figure} shows the 816 MHz emission detected by  MeerKAT around HESS~J1640$-$465.~The extent of the H.E.S.S. source is indicated by the white circle \citep{2014MNRAS.439.2828A}, its Fermi$-$LAT counterpart with the purple circle \citep{2021ApJ...912..158M}.~The associated SNR G338.3$-$0.0 \citep{2011_castellati} is outlined in teal, while the X-ray PWN  is shown as a blue ellipse \citep{2014ApJ...788..155G}. The position of the X-ray pulsar PSR J1640$-$4631 is indicated by the white cross \citep{2009ApJ...706.1269L}. Exact coordinates and extent of these sources are listed in Table \ref{tab:regions}. 

Within the SNR, we also detect a point source (PS; marked in green in Figure~\ref{fig:new_figure}), whose position is also listed in Table~\ref{tab:regions}, and was previously reported by \citet{2011_castellati}.~Their analysis of radio observations of this field suggested the flux density of this PS to be  $\sim6~{\rm mJy}$ and $\sim4.5~{\rm mJy}$ at 4800 and 8640 MHz, respectively. At 816 MHz, we measure a flux density of $S_\nu \sim 62 \pm 4.1$ mJy. A power-law fit to these three measurements yields a spectral index of $\alpha \sim -1.1$ ($S_\nu \propto \nu^{\alpha}$), which is somewhat flatter than typically observed for radio pulsars \citep{2018MNRAS.473.4436J}. This supports the earlier conclusion by \citet{2011_castellati} that this PS is unlikely to be a radio pulsar.

An examination of Figure~\ref{fig:new_figure} also reveals a region of enhanced diffuse radio emission near the geometric center of SNR G338.3$-$0.0.~This emission appears distinct from the bright shell-like structure characteristic of the remnant and suggests the presence of a separate central component to the radio emission of this source. To assess its significance, we first estimated the average noise-level in this image using 
the root-mean-squared (rms) intensity from a source-free region, obtaining $\sigma \approx 1.654~{\rm mJy}~{\rm beam}^{-1}$.  As shown in Figure~\ref{fig:contours}, the surface brightness of emission detected within SNR G338.3$-$0.0 is $\gtrsim5\sigma$ - a level sufficiently significant to warrant further investigation.

To quantify the morphology and spatial extent of this central emission, we fit the shape of the innermost 5$\sigma$ contour level,
with an ellipse -- implying this central diffuse radio emission has a semi-major and semi-minor axes of $\sim2^\prime\!.0$ and $\sim1^\prime\!.65$, respectively.~We adopt the properties of the fitted ellipse as the boundary of what we hereafter refer to as the interior radio diffuse emission region. The center of this region is given in Table~\ref{tab:regions} and 
marked by a white cross in Figure~\ref{fig:contours}, while its size is shown as a yellow ellipse in Figure~\ref{fig:new_figure}. While the centroids of the interior radio diffuse emission and the VHE $\gamma$-ray emission differ, the spatial overlap of this region with HESS~J1640$-$465, PSR~J1640$-$4631, and the \textit{NuSTAR} X-ray PWN suggests a possible physical connection between the emission detected across these different wavebands.

To determine the distribution of radio emission within this SNR, we constructed a radial brightness profile by averaging the emission within concentric circular annuli centered on this region, extending out to a radius of $7^\prime$, while excluding the contribution from the compact PS discussed above. The resulting profile, shown in the right panel of Figure~\ref{fig:radial_profile}, shows that the surface brightness peaks at the center of the remnant, and gradually decreases until the  semi-minor and semi-major axes of the interior diffuse emission derived above (black and purple dashed lines in Figure \ref{fig:radial_profile}), where it reaches the background level estimated in \S\ref{sec:background}.~Beyond this distance, the average surface brightness begins to rise until it peaks within the SNR shell (green dashed line; \citealt{2011_castellati}).~This deficit in  between the interior and shell of the SNR suggests distinct physical origins for these components, as will be discussed in \S\ref{sec:nature}.

\subsection{L Band Analysis} \label{sec:l_band}

Due to the higher frequency of L-band (856$-$1712 MHz), these observations have a better angular resolution ($\theta_{\rm res}$ $\sim3.\!\!^{\prime \prime}5$) than those in the UHF band ($\theta_{\rm res}$ $\sim6.\!\!^{\prime \prime}9$).  Unfortunately this comes at the expense of sensitivity to diffuse, extended emission.  To enhance the sensitivity of the resultant images to larger scales, we reimaged the data with weighting schemes that emphasize shorter baselines -- specifically using the \texttt{Briggs} weighting \citep{1995PhDT.......238B} with different robustness settings (-0.5, 0 \& +0.5) using \texttt{CASA} (\citealt{2007ASPC..376..127M}; \citealt{2022PASP..134k4501C}) to find an optimal balance between sensitivity and resolution that would accentuate the fainter diffuse emission from the PWN at higher frequencies. We used the task \texttt{tclean} to generate images at 1.4 GHz in \texttt{CASA}. Despite these adjustments, the emission from the PWN at this higher frequency is less noticable, making its difficult to accurate characterization its properties.~Additionally, when attempting to measure the background surface brightness, we consistently obtain negative flux values, indicative of missing flux due to incomplete sampling of the u-v plane at this higher frequency.  Consequently, we only place a lower limit on the PWN flux density at 1.4 GHz to be $S_{1284} \sim 60 \ {\rm mJy}$, following the procedure similar to that used for the UHF images described above.

\begin{figure}[htbp]
    \centering
    \includegraphics[width=1.0\linewidth]{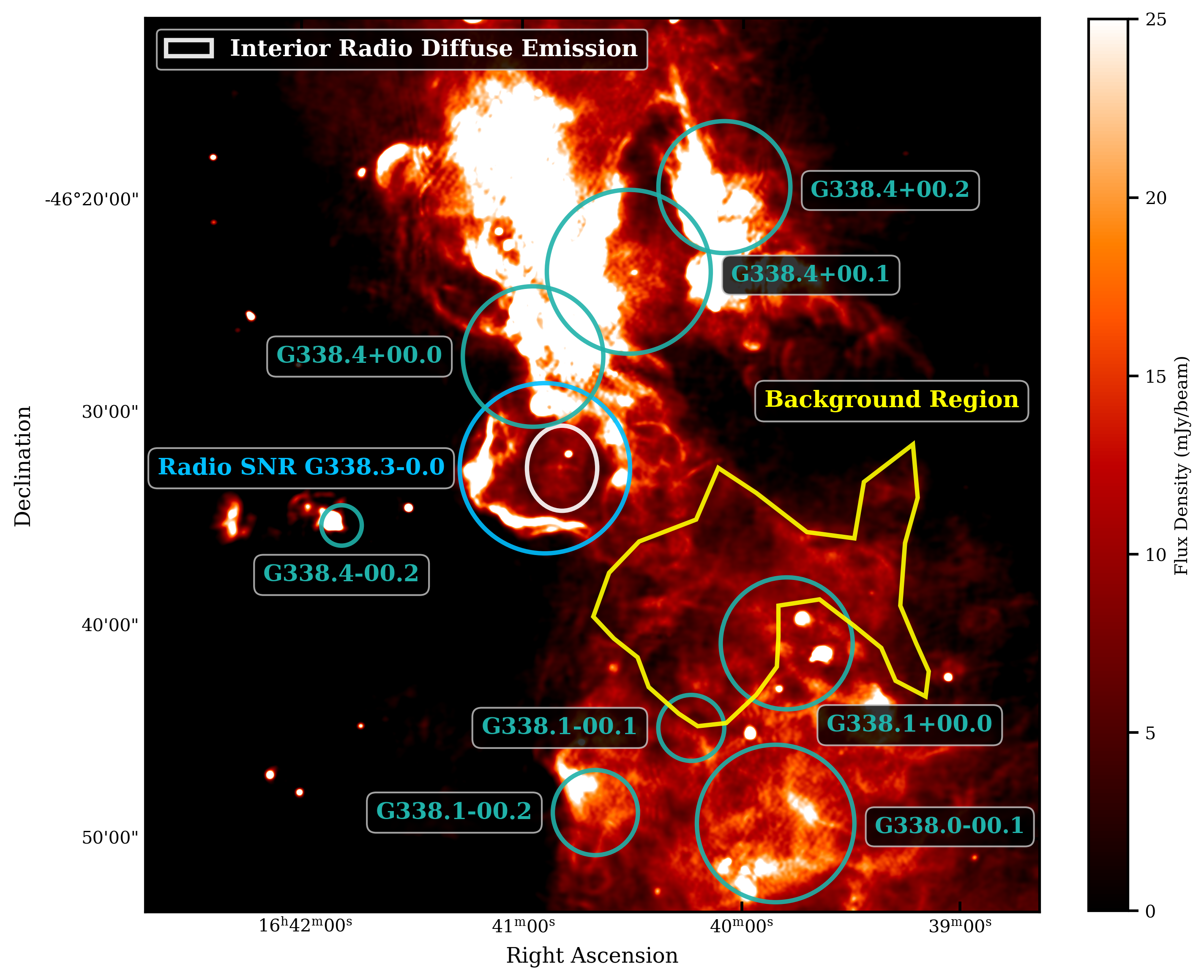}
    \caption{MeerKAT 816 MHz radio continuum image covering a $0.7^\circ$ field centered on SNR G338.3$-$0.0 (blue circle). Nearby catalogued thermal sources are outlined in teal, based on the radio catalog of galactic H\,\textsc{ii} regions compiled by \citet{2003A&A...397..213P}. The white ellipse marks the interior radio diffuse emission, while the yellow region denotes one of the selected background areas. The color scale is linear, with intensity in units of mJy\,beam$^{-1}$.}
    \label{fig:thermal_sources}
\end{figure}

\begin{figure*}[htbp]
        \centering
        \includegraphics[width=1.0\linewidth]{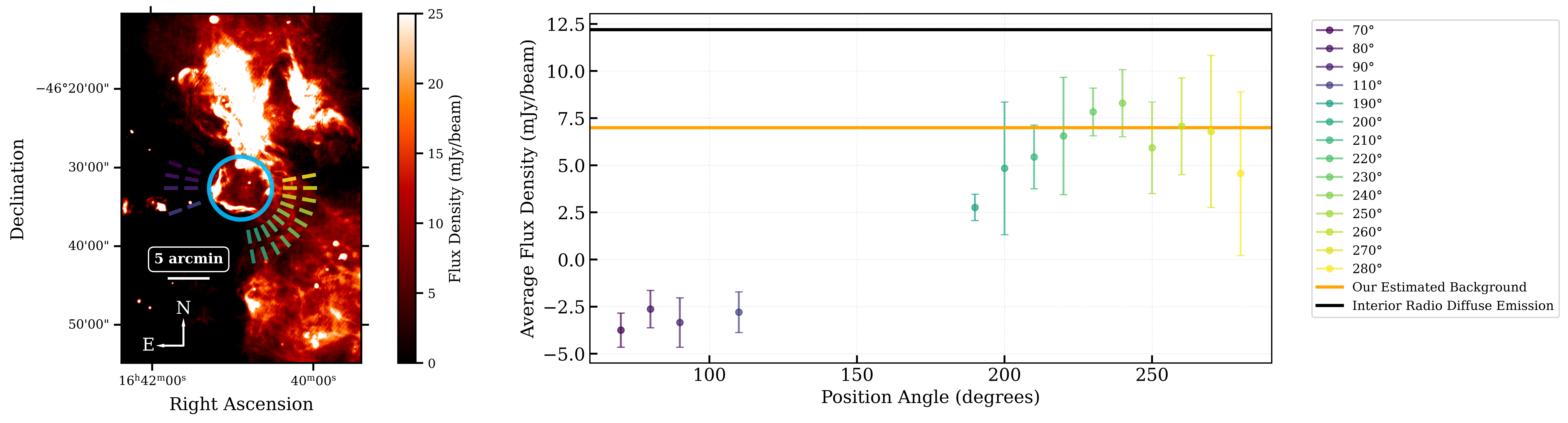}
\caption{\textit{Left panel:} FoV around SNR G338.3$-$0.0 (blue circle) showing the radial directions along which the flux density (in mJy\,beam$^{-1}$) was estimated. Each dashed line corresponds to a specific position angle and is color-coded to match its counterpart in the right panel. The color scale is linear, with intensity in units of mJy\,beam$^{-1}$ \textit{Right Panel:} Mean flux density along each radial direction plotted against position angle.~The estimated background level (See \S \ref{sec:background}) is indicated by a horizontal orange line, and the level of the interior diffuse emission is indicated by a horizontal black line. Error bars reflect the rms variation in the measurements. Angles are measured east of north. }
\label{fig:background_angle}
\end{figure*}

\section{Properties of The Interior Diffuse Emission}
\label{sec:interior_prop}

\subsection{Background Estimation}
\label{sec:background}

As mentioned in \S\ref{sec:UHF_band_analysis} and \S\ref{sec:l_band}, in both bands the images produced from the MeerKAT observation indicate the presence of diffuse radio emission inside SNR G338.3$-$0.0.  However, due to the location of this SNR and associated $\gamma$-ray emission in a crowded region of the Galactic plane (Figure \ref{fig:thermal_sources}), the detected emission within this SNR may include contributions from the diffuse Galactic background and/or unrelated astrophysical sources along the line of sight.~Determining the flux density, origin, and nature of the radio emission generated within this SNR requires estimating the surface brightness of background emission at this location.

Following the work of \citet{2011_castellati}, we extracted one-dimensional intensity profiles along multiple radial directions from the edge of the SNR.  The left panel of Figure \ref{fig:background_angle} displays the radio continuum map of the HESS J1640$-$465 field, where SNR G338.3$-$0.0 denoted by a blue circle, with multiple straight lines extending radially outward, each representing a direction along which the flux density was measured.~The right panel shows the mean flux density (in mJy~beam$^{-1}$) along each direction against the corresponding position angle.~The color of each point matches its directional cut, and the error bars reflect the rms variation in that direction.

As evident from Figure~\ref{fig:background_angle}, we observed notable variations in the average flux density across different directions, with marked differences between the eastern and western regions.~This variation suggests that the local environment around HESS J1640$-$465 is not uniform, and that a simple averaging of these selected regions might not yield the most reliable background estimate. 

Therefore, we estimate the intensity of the background emission within SNR G338.3$-$0.0 using the emission observed from a nearby ``source-free' region (e.g., \citealt{gelfand07, gelfand07b, gelfand13}). To do so, we choose a background region close to the SNR which:
\begin{enumerate}
    \item does not overlap with known astronomical sources in the field, 
    \item is likely to have comparable contamination from diffuse thermal radio emission as the center of this SNR, and
    \item sufficiently large that the average surface brightness $\bar{\Sigma}_{\rm bkg}$ is insignificantly impacted by small variations in the boundary.
\end{enumerate}

To satisfy the first criterion above, we require our background to exclude the H\,\textsc{ii} regions cataloged by \citet{2003A&A...397..213P}, whose positions are shown in Figure \ref{fig:thermal_sources}.~Since thermal radio sources also produce emission across the infrared (IR) band, to satisfy the second condition, we choose a background region with an IR surface brightness similar to what is observed from the interior of SNR G338.3$-$0.0.~To satisfy the third criterion, we require the background region to have the largest possible area that satisfies the first two criteria, and will estimate the uncertainty in $\bar{\Sigma}_{\rm bkg}$ by calculating this quantity for regions with slightly different boundaries.~A representative background region satisfying all selection criteria is highlighted in yellow in Figure~\ref{fig:thermal_sources}, with the rationale for its choice is detailed in \S \ref{sec:appendix_background}.


\begin{table}[tbp]
\centering
\caption{Values of the parameters used to calculate the significance of the PWN detection following equation \ref{eqn:delchi}}.
\begin{tabular}{cc}
\toprule
\toprule
Parameter & Value \\
\midrule
$\bar{\Sigma}_{\rm bkg}$  & 141.3~$\frac{\rm \mu Jy}{\rm pixel}$ \\
$\sigma_{\rm bkg}$ & 4.386~$\frac{\rm \mu Jy}{\rm pixel}$\\
\hline
$\bar{\Sigma}_{\rm int}$ & 164.4~$\frac{\rm \mu Jy}{\rm pixel}$\\
$\sigma_{\rm int}$ & $0.5654~\frac{\rm \mu Jy}{\rm pixel}$ \\
\bottomrule
\end{tabular}
\label{tab:sigma}
\end{table}

\subsection{Statistical Significance}
\label{sec:source_sigma}

Having detailed our methodology for the selection of background regions in \S \ref{sec:background} \& \S \ref{sec:appendix_background}, we can now determine the statistical significance $\Delta \chi_{\rm int}$ of the incresed radio surface brightness detected towards the center of SNR G338.3$-$0.0, first discussed in \S\ref{sec:UHF_band_analysis}.~We define $\Delta \chi_{\rm int}$ to be:
\begin{eqnarray}
    \label{eqn:delchi}
    \Delta \chi_{\rm int} & =  & \frac{\bar{\Sigma}_{\rm int} - \bar{\Sigma}_{\rm bkg}}{\sqrt{(\sigma_{\rm int})^2 + (\sigma_{\rm bkg})^2}} 
\end{eqnarray}
where $\bar{\Sigma}_{\rm int, bkg}$ are the average surface brightness within the interior and background regions, respectively, while $\sigma_{\rm int, bkg}$ are the uncertainties of these average surface brightnesses.~A detailed calculation of each term in equation \ref{eqn:delchi} is provided in \S \ref{sec:appendix_stat_sign} with the resulting values provided in Table \ref{tab:sigma}.~Substituting these values into Equation \ref{eqn:delchi}, we find that $\Delta \chi_{\rm int} \approx 5.22$ -- i.e., the difference between the average surface brightness of emission from the center of this SNR and the average surface brightness of the background emission in this field (\S\ref{sec:background}) is $\gtrsim5\times$ higher than the total uncertainty in these quantities.

\subsection{Flux Density Calculations}
\label{sec:source_flux}
\subsubsection{Interior Diffuse Emission Region}

As established in \S\ref{sec:source_sigma}, the surface brightness of radio emission within this SNR is significantly brighter than the background regions (as chosen using the criteria described in \S\ref{sec:background}) in this field.~However, the flux density within in the interior diffuse region identified in \S\ref{sec:UHF_band_analysis} includes emission from the background discussed in \S\ref{sec:background}, and the unrelated PS (\S\ref{sec:UHF_band_analysis}). Therefore, measuring the flux density of the radio-emitting plasma within the SNR requires determining the contribution of these different components to observed total emission.~Following the procedure outlined in \S ~\ref{sec:appendix_pwn_flux_density}, we subtract the contributions from the PS and the background to estimate the flux density of the interior diffuse emission.~Using this method, we obtain an average flux density of $\bar S_{\rm int} = 180 \pm 34$\,mJy for the interior diffuse emission region.

\subsubsection{SNR Shell Flux Density Calculations}
\label{sec:snr_flux_appendix}

The flux density of the SNR shell was estimated by accounting for contributions from both the background and the interior diffuse emission region (see \S \ref{sec:appendix_snr_flux_density} for details). Using two complementary methods, we derive a weighted average flux density of $\bar S_{\rm SNR}^{\rm Final} = 6.70 \pm 0.18$\,Jy, consistent with the 843 MHz flux density reported by \cite{2011_castellati} after subtracting thermal emission. The final flux densities of the interior diffuse emission region, the SNR shell, and the point source are summarized in Table~\ref{tab:flux_density_measurement}.

\begin{table}[tbp]
    \caption{Measurements of the 816 MHz Flux Density $S_{816}$ of SNR G338.3$-$0.0, the radio PS inside the SNR, and the interior radio diffuse emission region.}
    \centering
    \begin{tabular}{c|cc} 
        \toprule
        \toprule
 Frequency       & Region & Flux Density\\ 
        \midrule
        \multirow{3}{*}{816 MHz} & Radio PS & $61.6 \pm 1.1$ mJy \\ 
                             & SNR G338.3$-$0.0 & $6.70 \pm 0.18$ Jy \\
                             & Interior Diffuse Emission & $180 \pm 34$ mJy \\
        \bottomrule
    \end{tabular}
    \tablecomments{The reported flux densities are the average of multiple measurements. Detailed calculations can be found in \S \ref{sec:source_flux}.}
\label{tab:flux_density_measurement}%
\end{table}

\section{Nature of Interior Diffuse Emission} 
\label{sec:nature}

\subsection{Possible Interpretations}

The interior radio diffuse emission detected within SNR G338.3$-$0.0 could plausibly arise from one of three physical origins:

\begin{enumerate}
    \item Thermal bremsstrahlung from  overlapping H\,\textsc{ii}  regions
    \item Non-thermal synchrotron emission associated with the SNR shell
    \item Non-thermal synchrotron emission from a central PWN.
\end{enumerate}

Below, we explore each scenario in turn.

\subsubsection{Diffuse Thermal Emission from an Overlapping H\,\textsc{ii}  Region} 

The first possible explanation is that the radio emission originates from thermal bremsstrahlung associated with an overlapping H\,\textsc{ii} region.~However, several observational factors argue against this explanation. First, there is no corresponding  MIR/FIR emission detected at 24 $\mu$m, 70 $\mu$m, or longer wavelengths (see Figure~\ref{fig:pwn_IR_survey}), as would typically be expected from warm dust or ionized gas associated with H\,\textsc{ii} regions \citep{2006ApJ...638..157M}. Second, the emission lies outside the boundaries of  catalogued thermal sources in this field \citep{2003A&A...397..213P}, as shown in Figure~\ref{fig:thermal_sources}.~These considerations  disfavor a thermal origin.

\subsubsection{Non-thermal Synchrotron Emission from the SNR Shell}

An alternative explanation is that the interior diffuse emission arises from non-thermal synchrotron radiation from the SNR shell. However, this scenario is less plausible, as SNRs typically exhibit a shell-like morphology characterized by surface brightness that decreases primarily monotonically toward the center, as observed in other SNRs such as Kes 17 (SNR G304.6+0.1; \citealt{kes17_gelfand}) and SNR G46.8$–$0.3 \citep{2022A&A...664A..89S}. In contrast, the radial profile plot shown in Figure \ref{fig:radial_profile} deviates markedly from this expectation.~Instead of a smooth decrease from the shell inward, the profile exhibits a pronounced central peak associated with the interior diffuse emission, followed by a decrease in brightness before rising again at the shell boundary.~This dip between the central emission and the shell suggests that the two components are physically distinct.

\subsubsection{Non-thermal Emission from a PWN}

With the previous two scenarios ruled out, the most likely interpretation is that the interior radio diffuse emission arises from a radio PWN within SNR G338.3$-$0.0.~Although a measurement of the spectral index is typically used to distinguish between thermal and non-thermal radio emission, the current data do not permit such an estimate for the interior diffuse emission in SNR G338.3$–$0.0.~The incomplete \textit{u$–$v} coverage at 1.4 GHz prevents a reliable estimate of the flux density, making it difficult to derive a meaningful spectral index.~While this limitation prevents a definitive confirmation of the synchrotron nature of the emission, similar reasoning has been used in other objects (e.g., PSR J0855$-$4644; \citealt{2018MNRAS.477L..66M}) to rule out a thermal origin and attribute the centrally peaked radio emission to a PWN.

\subsection{Implications of a Possible PWN Origin}

Looking at Figure \ref{fig:new_figure}, the interior radio diffuse emission region lies entirely within the GeV/TeV $\gamma$-ray emission region, as detected by both \textit{Fermi}-LAT and H.E.S.S, and partially overlaps with the X-ray PWN seen by \textit{NuSTAR} and the X-ray pulsar, with the peak of the radio emission being offset from both the X-ray emission peak and the pulsar position. Such offsets between radio and X-ray emission are not unusual in PWNe. In fact, the radio and X-ray morphologies are often not well correlated, and cases of even anti-correlation have been reported, such as the PWN associated with the Vela pulsar \citep{2003MNRAS.343..116D} and G319.9$-$0.7 \citep{2010ApJ...712..596N}.

Despite the general spatial association between the X-ray PWN and the interior radio diffuse emission region, they have different axes of symmetry (Figure \ref{fig:new_figure}). Such mismatches and disrupted morphologies are a hallmark of a late evolutionary displacement of the PWN by the SNR reverse shock (RS).~As the RS propagates asymmetrically through the inhomogeneous interior of a core-collapse SNR, it can compress and displace/distort the PWN, leading to the observed offsets and complex morphologies across multiple wavelengths \citep{2001ApJ...563..806B, 2004A&A...420..937V, 2006ARA&A..44...17G}. Such a disrupted morphology is not unique to HESS J1640$-$465 and has been observed in other evolved PWNe systems.~For example, G327.1$-$1.1 exhibits a radio "relic" lobe displaced from its X-ray head \citep{2009ApJ...691..895T, 2015ApJ...808..100T, 2022ApJ...940..143E}, while the composite SNR MSH~15$-$56 shows a cometary radio structure trailing behind its pulsar, with X-ray diffuse emission beyond the boundaries of the radio PWN \citep{2015ApJ...808..100T, 2017ApJ...851..128T}.~A similar  morphology has also been seen in G328.4+0.2 \citep{2007ApJ...663..468G}.~These morphologies reflect the different spatial distribution of the electron populations that dominates at different wavelengths:~extended radio emission arises from older, lower-energy electrons that have diffused outward, while the compact X-ray nebula is powered by recently injected, high-energy electrons confined closer to the pulsar due to rapid synchrotron cooling \citep{2006ARA&A..44...17G, 2022hxga.book...61M}. Any additional displacement due to the pulsar's motion can further distort the nebular morphology. Although direct measurements of PSR~J1640$-$4631's proper motion are not yet available, \citet{2016ApJ...819L..16A} reported a braking index of $n \simeq 3.15 \pm 0.03$, which is unusually high for pulsars.~This may indicate a potential misalignment between the pulsar's magnetic and rotational axes, which could give rise to proper motion, as suggested by \cite{2013Sci...342..598L}. However, further observational constraints are required to confirm this.

Therefore, the observed radio--X-ray offset and differing orientations inside the SNR G338.3$-$0.0 are naturally explained if the PWN has been ``crushed'' and swept off-center by a reverse shock interaction.~This evolutionary scenario is consistent with earlier modeling by \citet{2021ApJ...912..158M} and \citet{2023ApJ...946...40A}, both of which argue that a reverse shock encounter has already occurred in this system.~Therefore, the interior radio diffuse emission inside SNR G338.3$-$0.0 is best interpreted as the radio counterpart of the X-ray PWN.


\begin{figure*}[tbhp]
    \centering
    \begin{minipage}{0.45\textwidth}
        \centering
        \includegraphics[width=\linewidth]{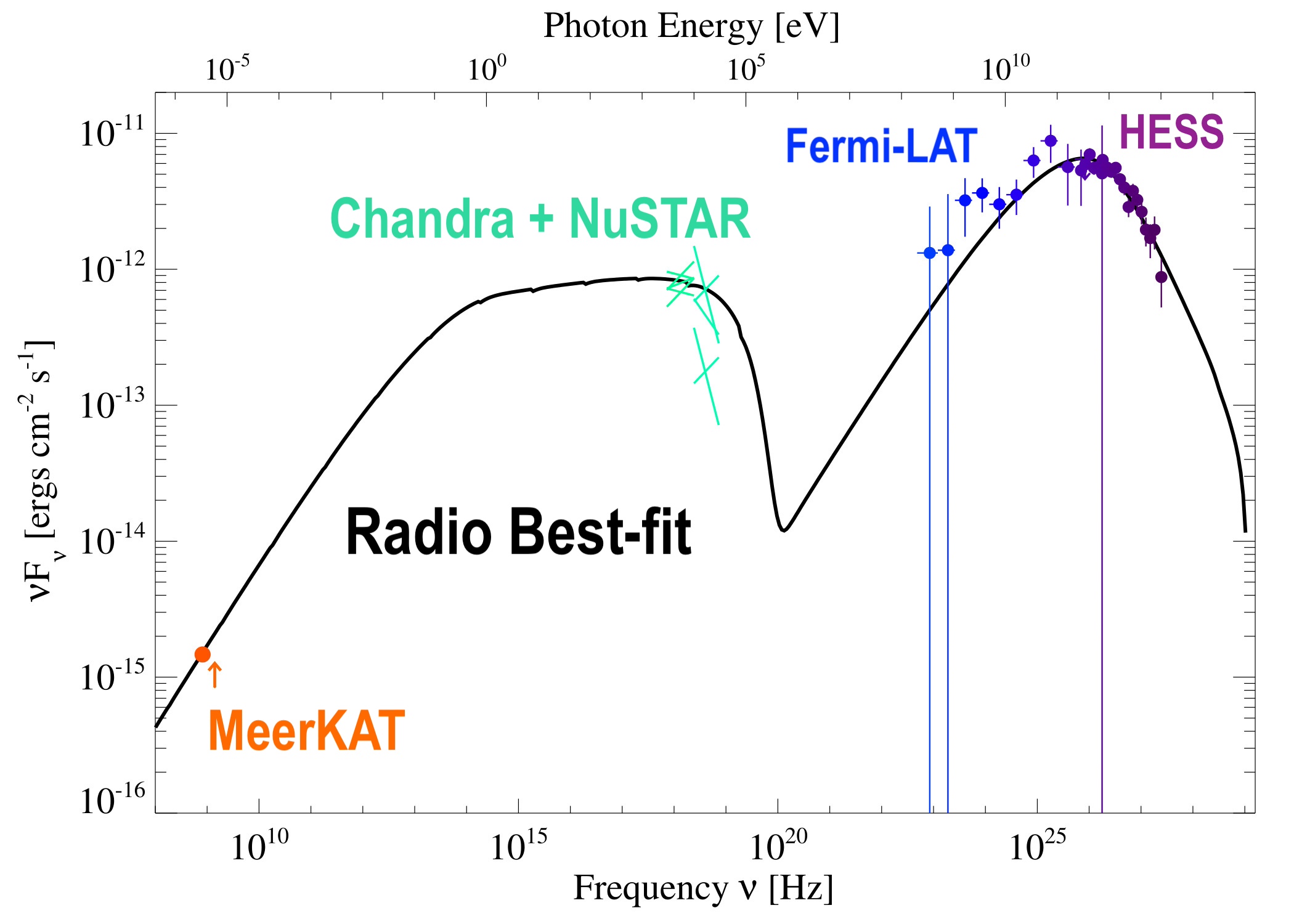}
        \caption*{(a)}
    \end{minipage}
    \hfill
    \begin{minipage}{0.45\textwidth}
        \centering
        \includegraphics[width=\linewidth]{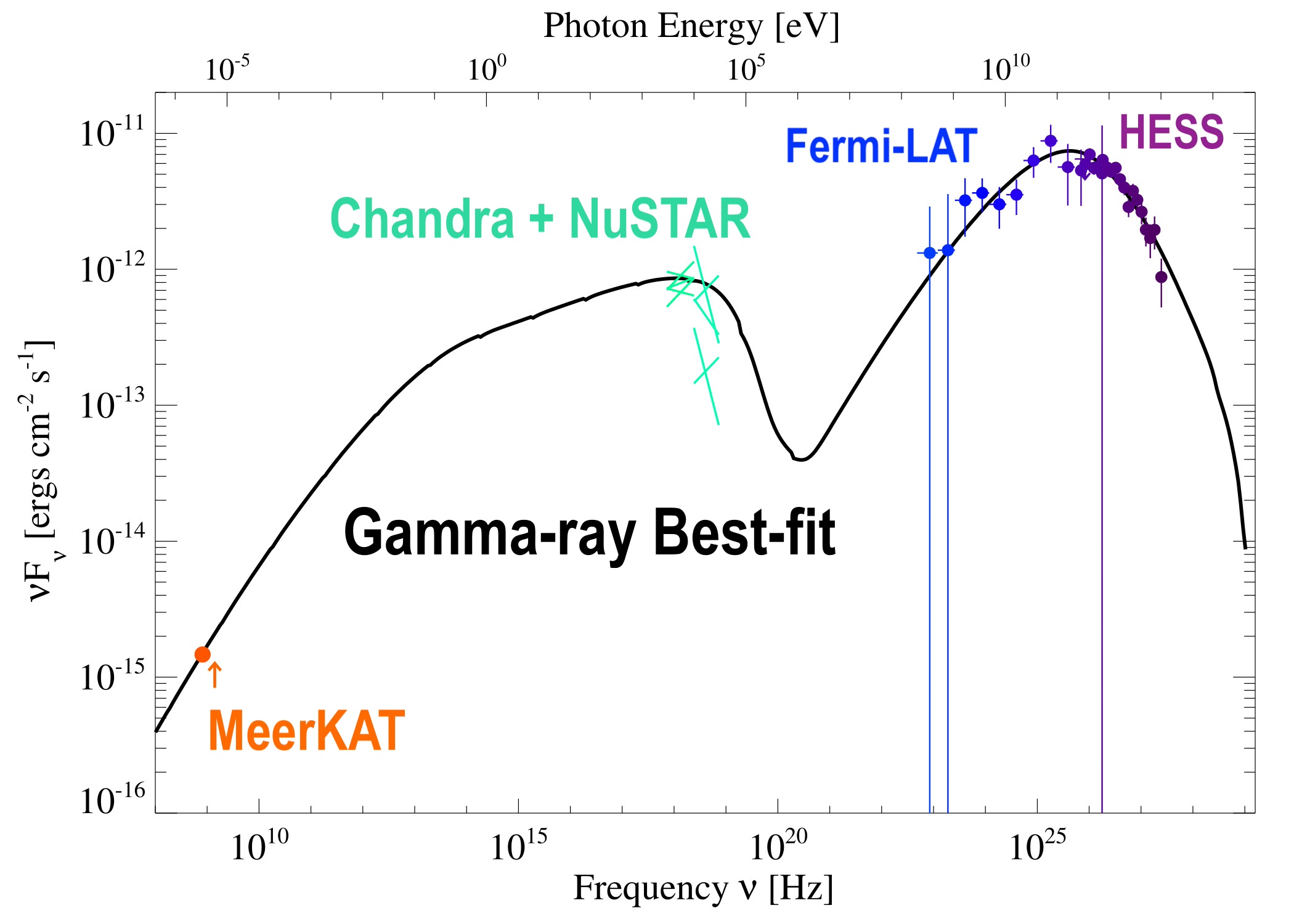}
        \caption*{(b)}
    \end{minipage}

    \caption{Top panel: SED model fit for HESS J1640$-$465, for two different PWN sizes: (a) the smaller radio size and (b) the larger $\gamma$-ray size. The best-fit model parameters are listed in Table \ref{tab:pwn_modelling}.}
    \label{fig:SED}
\end{figure*}

\begin{table*}
    \centering
    \caption{Our best-fit model parameters for HESS J1640$-$465 using two different sizes for the PWN along with previous results by \citet{2023ApJ...946...40A}}
    \label{tab:pwn_modelling}
    \renewcommand{\arraystretch}{1.3}
    \footnotesize
    \begin{tabular}{c|c|c|c}
        \toprule
        \toprule
        Model Parameter & Radio Size ($\theta_{\text{PWN}} \sim 1\farcm83$)
 & $\gamma$-ray Size ($\theta_{\text{PWN}} \sim 3\farcm3$) & \cite{2023ApJ...946...40A}\\
        \midrule
        SN Explosion Energy $E_{\text{sn}}$ & $5.6 \times 10^{51}$ erg  & $1.0 \times 10^{51}$ erg  &$1.4 \times 10^{51}$ erg   \\
        SN Ejecta Mass $M_{\text{ej}}$  & $9.5~M_{\odot}$ &$9.0~M_{\odot}$ & $10.0~M_{\odot}$  \\
        ISM Density $n_{\text{ism}}$ & $0.36~\mathrm{cm}^{-3}$& $0.01~\mathrm{cm}^{-3}$ & $0.03~\mathrm{cm}^{-3}$  \\
        Pulsar Spin-down Timescale $\tau_{\text{sd}}$ &12.6 yr & 5.0 yr & 4.76 yr \\
        Wind Magnetization $\eta_{B}$ & 0.0056  & 0.1321 & 0.07   \\
        Min. Injected Particle Energy $E_{\text{min}}$ & 1.0 GeV &  2 GeV & 1 GeV\\
        Break Injected Particle Energy $E_{\text{break}}$ & 1.38 TeV  & 1.33 TeV & 1 TeV\\
        Max. Injected Particle Energy $E_{\text{max}}$ & 0.73 PeV & 1.29 PeV&  1.24 PeV  \\
        Low Injected Particle Index $p_{1}$ &  1.71& 1.62 & 1.44 \\
        High Injected Particle Index $p_{2}$ &  2.87 & 2.72 & 2.68\\
        Temperature of Added Photon Fields $T_{\text{ic}}$ & 308 K & 360 K, 10454 K & 354 K, 16108 K\\
        Energy Density of Added Photon Fields $U_{\text{ic}}$ & $12~\frac{\mathrm{eV}}{\mathrm{cm}^{3}}$ &  $44~\frac{\mathrm{eV}}{\mathrm{cm}^{3}}$, $29~\frac{\mathrm{eV}}{\mathrm{cm}^{3}}$ & $580~\frac{\mathrm{eV}}{\mathrm{cm}^{3}}$, $170~\frac{\mathrm{eV}}{\mathrm{cm}^{3}}$ \\
        Distance $d$ & 11.4 kpc & 11.5 kpc & 11.9 kpc\\
                \hline
        $\chi^2_{\nu}$ / degrees of freedom & 1.21 / 24  & 1.5 / 22 & 1.36 / 22  \\
        \bottomrule
        \bottomrule
        \multicolumn{4}{c}{Select Output Properties} \\
       $\theta_{PWN}$ & $1\farcm92$  &  $3\farcm54$ &  $3\farcm45$  \\
       True Age $t_{age}$ & 3086 Yr & 3093 yr & 3093 yr \\
       Reverse Shock Time $t_{rs}$ & t=1905 yr &  t=2971 yr &  Yes$^*$ \\
      Current Magnetic Field Strength $B_{PWN}$ &$3.78~\mu\mathrm{G}$ & $5.5~\mu\mathrm{G}$ & $4.1~\mu\mathrm{G}$\\
        \bottomrule 
    \end{tabular}
    \tablecomments{ $t_{rs}$ is the age at which the PWN colides with the reverse shock. \\
    * ``Yes'' indicates that a reverse-shock interaction occurred, though the exact time was not reported in the cited paper.}
\end{table*}

\section{Modelling}
\label{sec:modelling}

Assuming the interior diffuse radio emission represents the radio counterpart to the X-ray PWN, this detection provides new observational constraints on both the size and flux density of the PWN in SNR G338.3$-$0.0 across multiple wavelengths.~As summarized in Table~\ref{tab:regions}, the radio, X-ray, and $\gamma$-ray observations show differences in the apparent size of the nebula. Such discrepancies in size across different bands have been observed for other TeV PWNe such as the Boomerang \citep{2024ApJ...960...75P}, the Dragonfly \citep{2023ApJ...954....9W}, and the Eel \citep{2022ApJ...930..148B}.~This can be understood in terms of the differing cooling timescales of the underlying electron populations responsible for emission at these energies \citep{2006ARA&A..44...17G, 2022hxga.book...61M}.

To reproduce the broadband dynamical and spectral properties of HESS J1640$-$465-including the additional constraints provided by the new radio measurement at 816 MHz and the size of the radio PWN, we use a time-dependent one-zone model developed by \citet{2009ApJ...703.2051G}. This model provides a self-consistent treatment of the PWN interaction with its host SNR, allowing us to infer key physical parameters of the system, including the supernova explosion energy, ejecta mass, and the pulsar’s initial spin properties, which are otherwise difficult to constrain observationally.~A comprehensive description of the model, along with the methodology for selecting the input parameters and fitting the observed PWN properties  using the Metropolis Markov Chain Monte Carlo (MCMC) algorithm  to match model predictions is provided by \citet{2020ApJ...904...32H} and references therein.

Because our one‑zone model treats the same e$^{\pm}$ population as the source of both synchrotron (radio \& X‑ray) and inverse Compton ($\gamma$-ray) emission, it implicitly assumes a single spatial region/ a common size for all energies.~Given the uncertainties in the nebular extent across wavelengths and the inherent limitations of spherical, one-zone models in capturing the energy-dependent morphology of PWNe, we adopt a conservative modeling approach. Specifically, we explore two scenarios:~(1) a “small” PWN with the radio size estimated in this work ($\theta_{\text{PWN}} \sim 1\farcm83$ \footnote{Computed as the average of the semi-major and semi-minor axes of the interior radio diffuse emission region listed in Table~\ref{tab:regions}.}), and (2) a “larger” PWN based on the $\gamma$-ray size ($\theta_{\text{PWN}} \sim 3\farcm3$) used in previous studies \citep{2023ApJ...946...40A, 2021ApJ...912..158M}.~These cases allow us to explore the plausible physical conditions of the system while acknowledging the model’s underlying assumptions. This approach has been used when modelling systems with discrepant measurements (e.g, the Eel; \citealt{2022ApJ...930..148B}).~In both scenarios, we include the new measurement of the radio flux density at 816 MHz and the upper limit at 1.4 GHz as additional constraints on the broadband spectral energy distribution (SED) of HESS J1640$-$465.

In our modelling, we assume a spin-down luminosity of  $\dot{E} = 3.72 \times 10^{36}~\mathrm{erg~s}^{-1}$ and a characteristic age of  $t_{\text{ch}} = 3350$ yr.~For consistency with earlier modeling efforts \citep{2023ApJ...946...40A, 2021ApJ...912..158M}, we fixed the pulsar braking index to $p = 3.15$, as measured by  \cite{2016ApJ...819L..16A}. To assess how the addition of the new radio measurements affects the inferred physical properties of the system, we directly compare our results for the $\gamma$-ray size scenario to those from \citet{2023ApJ...946...40A}. Figure~\ref{fig:SED} shows the best-fit SED for both cases considered in this work, and Table~\ref{tab:pwn_modelling} summarizes our best-fit model parameters.

\begin{figure*}[tbhp]
    \centering
    \begin{minipage}{0.45\textwidth}
        \centering
        \includegraphics[width=\linewidth]{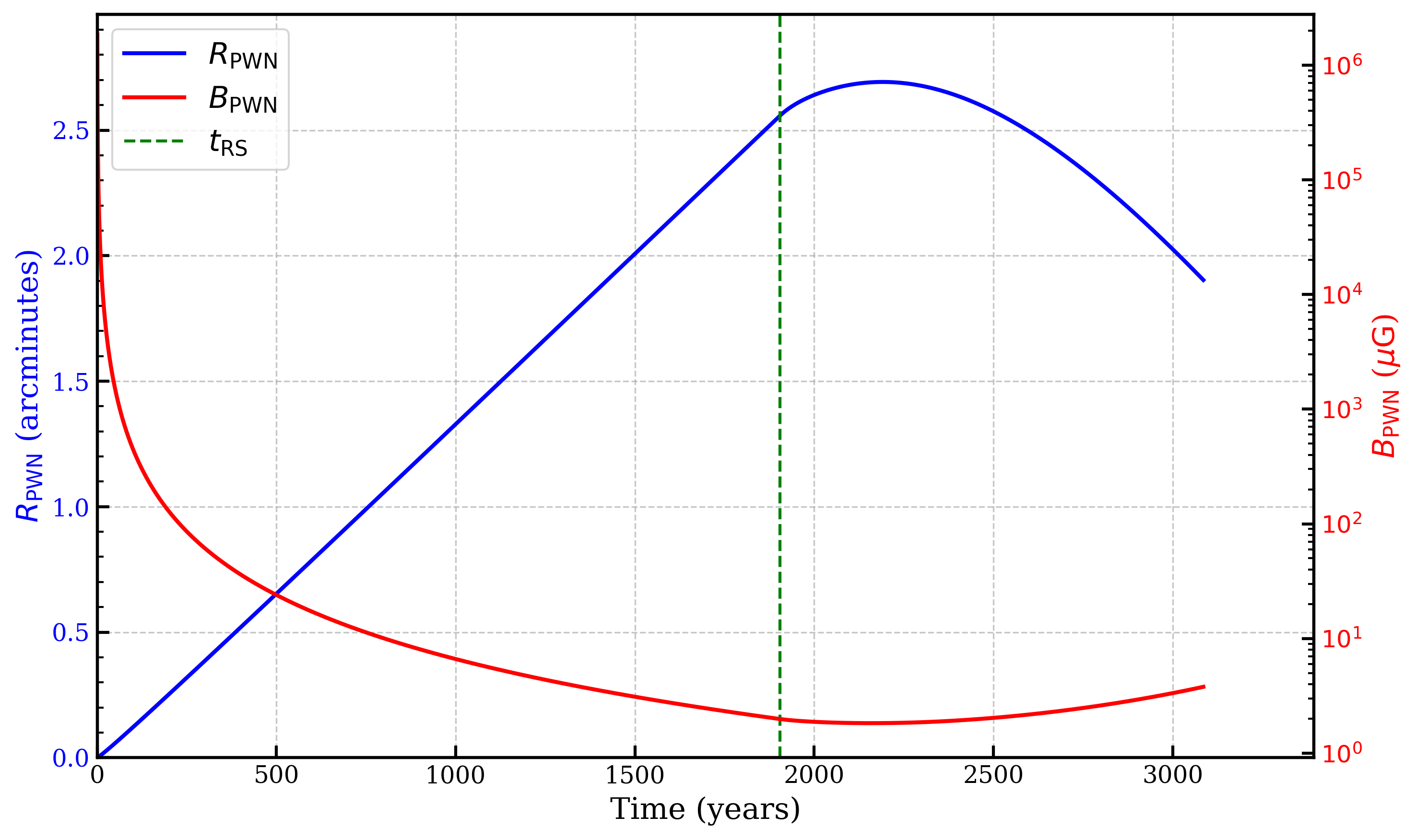}
        \caption*{(a) Radio Size Scenario}
    \end{minipage}
    \hfill
    \begin{minipage}{0.45\textwidth}
        \centering
        \includegraphics[width=\linewidth]{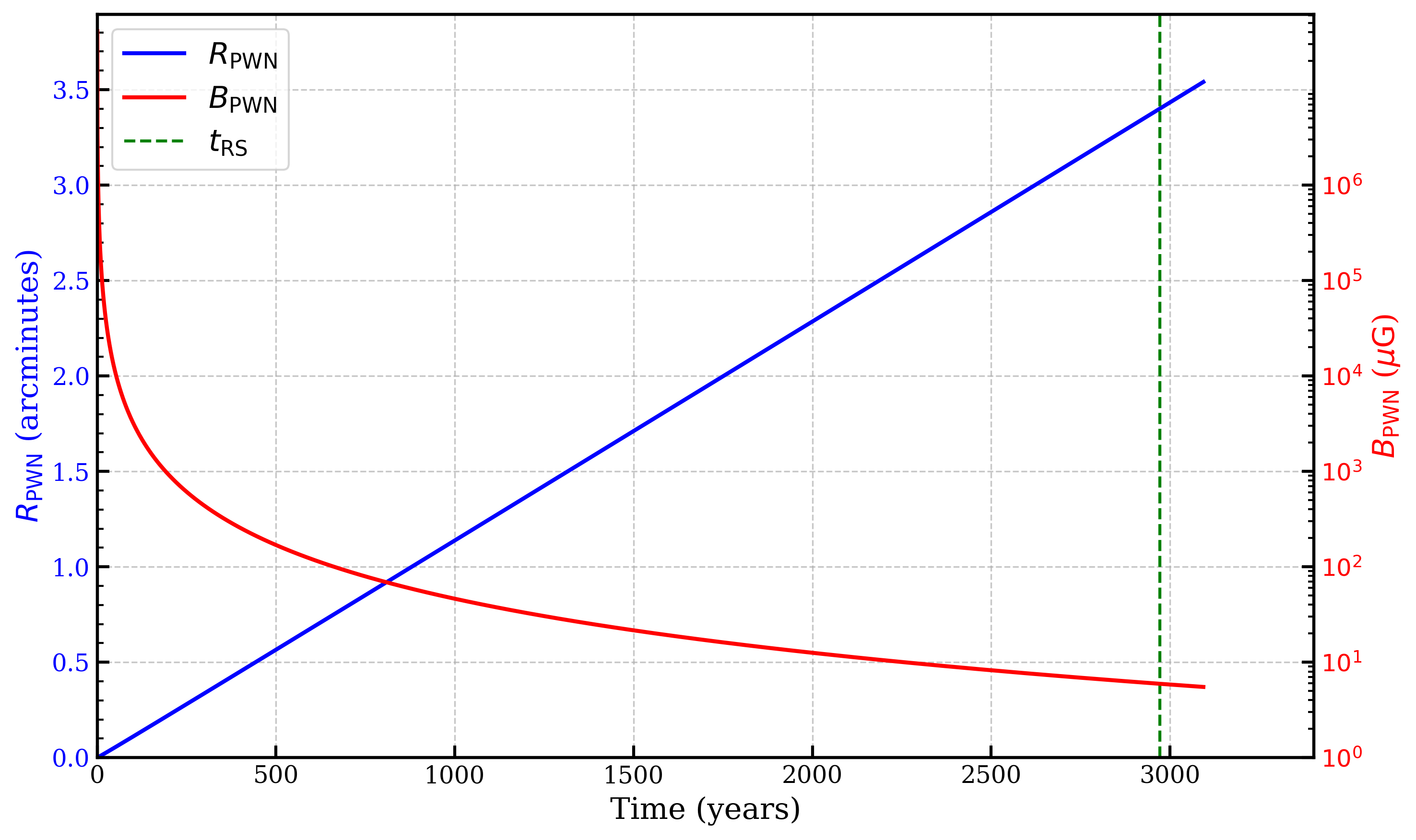}
        \caption*{(b) $\gamma$-ray size Scenario}
    \end{minipage}

    \caption{Evolution of the PWN radius ($R_{\mathrm{PWN}}$, blue solid line) and the PWN magnetic field ($B_{\mathrm{PWN}}$, red solid line) over the model’s true age ($t_{\mathrm{age}} \sim 3100$~yr) for the two scenarios. The green dashed lines mark the time of collision between the PWN and the SNR reverse shock, occurring at $t_{RS} \sim 1900$~yr for the radio size and $t_{RS} \sim 3000$~yr for the $\gamma$-ray size.}
    \label{fig:b_field}
\end{figure*}

\subsection{Inferred Properties from the Modelling}

Both scenarios considered in this work successfully reproduce the observed broadband properties of HESS\,J1640$-$465 within a purely PWN leptonic framework.~In both cases, the electron population must reach maximum energies exceeding $0.1\,\mathrm{PeV}$ to account for the TeV emission, supporting the interpretation of HESS\,J1640$-$465 as a Galactic PeVatron candidate. Moreover, the PWN is currently interacting with the reverse shock, an evolutionary phase usually linked to particles escaping into the ISM, suggesting that HESS~J1640$-$465 may be a source of Galactic PeV e$^{\pm}$.~In this section, we highlight some key results of our multi-wavelength modeling and their implications for the source’s physical properties and evolutionary state.

\subsubsection{Comparison with Previous $\gamma$‑ray Size Results}

Because our ``large'' PWN scenario adopts the same $\gamma$-ray size used by \citet{2023ApJ...946...40A}, we observe only minor differences in our fitted parameters compared to theirs. This isn't surprising given that our estimated flux density ($S_{816} = 180 \pm 34$\,mJy) is not incompatible with the predicted value of $S_{816} \sim 100$\,mJy.~The primary difference between the two fits lies in the background photon fields:~whereas \citet{2023ApJ...946...40A} required two components with energy densities of $U_{\text{ic,1}}$ $\sim 170$\,eV\,cm$^{-3}$ and  $U_{\text{ic,2}}$ $\sim 580$\,eV\,cm$^{-3}$, our updated fit prefers somewhat lower values (  $U_{\text{ic,1}}$ $\sim 29$\,eV\,cm$^{-3}$ and  $U_{\text{ic,2}}$ $\sim 44$\,eV\,cm$^{-3}$), driven by the addition of the new radio measurement at 816 MHz. 

\subsection{Comparison between the Radio and $\gamma$‑ray–Size Scenarios}

Adopting the $\gamma$-ray size ($\theta_{\rm PWN} = 3\farcm3$) yields \(E_{\mathrm{SN}} \sim 1.0 \times 10^{51}\) erg,
\(M_{\rm ej} \sim 9\,M_\odot\) and \(n_{\rm ism} \approx 0.01\,\mathrm{cm}^{-3}\), in excellent agreement with the results of \citet{2023ApJ...946...40A}.~In contrast, reproducing the smaller, radio-derived size ($\theta_{\rm PWN} = 1\farcm83$) at the same system age ($t_{age} \sim 3100$\,yr) requires a higher explosion energy (\(E_{\mathrm{SN}} \sim 5.6 \times 10^{51}\)) and a much denser medium \(n_{\rm ism} \sim 0.36\,\mathrm{cm}^{-3}\) to reproduce the same present‐day radius, while the ejecta mass remains \(M_{\rm ej} \sim 9.5\,M_\odot\).~In the radio-size scenario, these differences lead to an earlier interaction with the reverse shock at $t_{\rm RS} \sim 1900$\,yr, which compresses the nebula, reduces its size,and amplifies its magnetic field to $B_{\rm field} \sim 3.8~\mu\mathrm{G}$ by the present system age (see Figure \ref{fig:bfield_radio}).

By comparison, the lower ISM density and explosion energy in the $\gamma$-ray–size scenario delay the reverse shock interaction to approximately the current system age of $t \sim 3100$\,yr. This allows the PWN to continue expanding to the larger size observed at higher $\gamma$-ray energies, resulting in a present-day magnetic field of $B_{\rm PWN} \sim 5.5~\mu\mathrm{G}$ (see Figure~\ref{fig:bfield_gamma}).~This comparison demonstrates how the assumed PWN size directly influences the inferred properties, and the evolutionary stage of the system.

In both scenarios, the IC emission responsible for the observed $\gamma$-rays requires photon fields in addition to the CMB. For the radio-size scenario, we require a single photon field at T $\sim 300$ K with an energy density of $U_{\text{ic}} \sim 12$ eV cm$^{-3}$. In the $\gamma$-ray–size scenario, we require two fields: one at T$\sim 360$ K, with $U_{\text{ic,1}} \sim 44$ eV cm$^{-3}$ and another at T $\sim 10400$ K and $U_{\text{ic,2}} \sim 29$ eV cm$^{-3}$. While they are different from each other, both have energy densities higher than the standard interstellar radiation fields (ISRF) predicted by Galactic Cosmic Ray Propagation (GALPROP) code/framework \citep{galprop} for this region of the Galaxy (Galactocentric radius $\sim 6$ kpc), where typical values are $U_{\text{ic}} \sim 0.4$ eV cm$^{-3}$ for the FIR at T $\sim 26.5$ K, and $U_{\text{ic}} \sim 0.18$ eV cm$^{-3}$ for the NIR at T $\sim 2800$ K. Such higher and more extreme energy densities are not unexpected given the complex and nearby bright IR sources, as shown in Figures  \ref{fig:thermal_sources}, \ref{fig:background_IR_survey} \& \ref{fig:pwn_IR_survey}. Additionally, as noted by \cite{2021ApJ...912..158M} and \cite{2014ApJ...788..155G}, the young massive stellar cluster Mc81 \citep{2012MNRAS.419.1860D}, lies at a comparable distance ($d = 11 \pm 2$ kpc), consistent with the distance inferred from our PWN modeling ($d \sim 12$ kpc; Table~\ref{tab:pwn_modelling}).~Given its projected separation from HESS J1640$-$465 of $\sim 25$ pc \citep{2014ApJ...788..155G}, Mc81 is likely close enough to contribute significantly to the enhanced photon fields required to explain the observed $\gamma$-ray emission through ICS.

\section{Summary \& Future Steps } \label{sec:summary}

With new  observations using the MeerKAT radio interferometer, we have detected a statistically significant diffuse radio emission within SNR G338.3$-$0.0 ($S_{816} = 180 \pm 34  \ {\rm mJy}$; $S_{816} \gtrsim 60  \ {\rm mJy}$) associated with the TeV $\gamma$-ray source HESS J1640$-$465. This region, delineated by the 5$\sigma$ contour (Figure \ref{fig:contours}), overlaps the GeV/TeV $\gamma$-ray emission regions, the X-ray PWN, and PSR J1640$-$4631 (Figure \ref{fig:new_figure}).~An elliptical fit to this region yields a PWN radius of $\theta_{\rm PWN} \sim 1'.8$.

A lack of MIR and FIR counterparts (Figures \ref{fig:background_IR_survey} \& \ref{fig:pwn_IR_survey}), together with the absence of overlapping, catalogued H\,\textsc{ii} regions (Figure \ref{fig:thermal_sources}), argues against thermal bremsstrahlung as the emission mechanism.~Likewise, the centrally peaked radial brightness profile (Figure \ref{fig:radial_profile}) is incompatible with the monotonic decline expected for shell‑synchrotron emission, disfavoring an SNR shell. With thermal and shell scenarios excluded, the most plausable explanation is that we have detected the radio counterpart of the X-ray PWN powered by PSR J1640$–$4631. The spatial coincidence—offset from both the X‑ray peak and the pulsar position is characteristic of reverse‑shock–crushed PWNe, and mirrors morphologies seen in other evolved composite SNRs.

To explore the implications of this interpretation, we incorporated our new radio flux and size measurements into a one‑zone, time‑dependent evolutionary model. Two size scenarios were explored; a “small” (radio) and “large” ($\gamma$‑ray), each reproducing the broadband SED under a purely leptonic framework. Both scenarios require electron energies exceeding $0.1\,\mathrm{PeV}$ to reproduce the TeV emission, indicating that HESS\,J1640$-$465 is a Galactic PeVatron candidate.

In terms of the next steps, more extensive modelling could be performed to explore certain degenerecies between the model parameters (e.g, $\tau_{sd}$ \& p), similar to what was done for PWN G54.1+0.3 \citep{2015ApJ...807...30G}.~This will help constrain the properties of the supernova explosion and neutron star responsible for the luminous VHE $\gamma$-ray source, and the spectrum of the e$^\pm$ it contains. This is important for understanding the contribution of PWNe to Galactic PeV e$^\pm$.

\section*{Orcid IDs}

\begin{tabbing}
\hspace{3cm} \= \hspace{2cm} \= \kill
\text{Moaz Abdelmaguid:} \> \orcidlink{0000-0002-4441-7081}\href{https://orcid.org/0000-0002-4441-7081}{0000-0002-4441-7081} \\
\text{Joseph D. Gelfand:} \> \orcidlink{0000-0003-4679-1058}\href{https://orcid.org/0000-0003-4679-1058}{0000-0003-4679-1058} \\
\text{J. A. J. Alford:} \> \orcidlink{0000-0002-2312-8539}\href{https://orcid.org/0000-0002-2312-8539}{0000-0002-2312-8539}
\end{tabbing}

\section{Acknowledgments}
M.A. is supported by a graduate research assistanship at NYU Abu Dhabi, which is funded by the Executive Affairs Authority of the Emirate of Abu Dhabi, as adminstrated by Tamkeen.~The research of J.D.G is funded by NYUAD grant AD 022.~Both M.A. and J.D.G. received support form the NYU Abu Dhabi Research Institute grant to the CASS. The MeerKAT radio telescope is operated by the South African
Radio Astronomy Observatory, which is a facility of the National Research Foundation, an agency of the South Africa
Department of Science and Innovation. \\

\noindent \emph{Facilities}: MeerKAT, $\chandra$ \\
\emph{Software}: Miriad \citep{1995ASPC...77..433S} \& CASA (\citealt{2007ASPC..376..127M}; \citealt{2022PASP..134k4501C})

\appendix
\section{Flux Density Calculations}
\label{sec:appendix}
\renewcommand{\thefigure}{A.\arabic{figure}}
\setcounter{figure}{0}
\renewcommand{\thetable}{A.\arabic{table}}
\setcounter{table}{0}

In order to accurately quantify the diffuse emission within SNR G338.3$-$0.0, it is necessary to estimate the contribution from unrelated background sources and diffuse Galactic emission.~This appendix describes the rationale for selecting the background regions and the steps used to calculate the flux densities of the background, the PS, the SNR shell, and the radio interior diffuse emission region.

\subsection{Background Choice} 
\label{sec:appendix_background}

\begin{figure*}[htbp]
        \centering
        \includegraphics[width=1.0\linewidth]{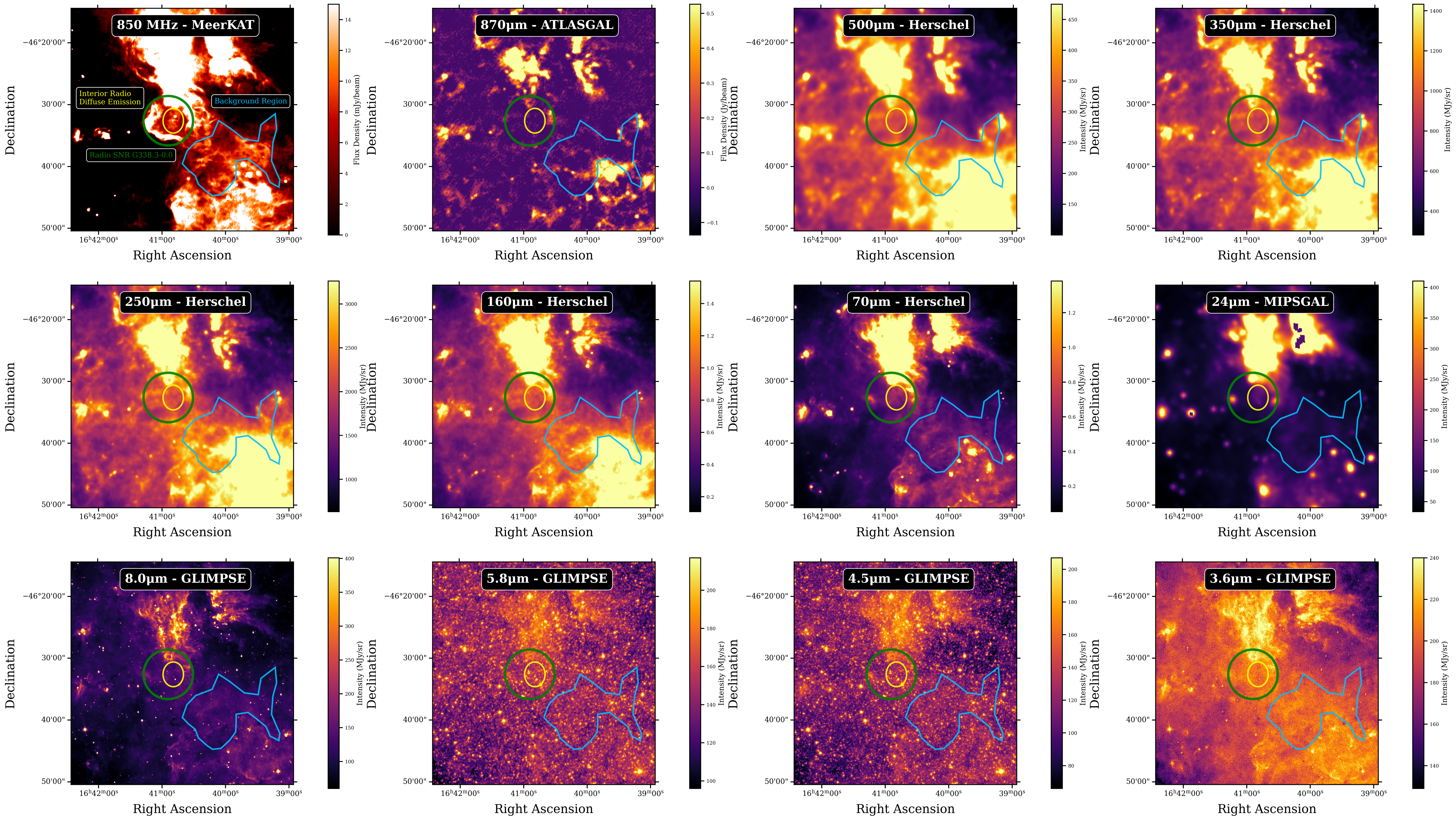}
\caption{ Multi-wavelength comparison of the region surrounding SNR G338.3$-$0.0 to ensure a representative background choice.~The leftmost panel shows the MeerKAT 816 MHz image, with the interior radio diffuse emission (yellow ellipse), one of the selected background regions (blue), and SNR  G338.3$-$0.0 (green circle). The remaining panels display the corresponding FoV from mid-infrared (MIR) and far-infrared (FIR) surveys, including ATLASGAL at 870~$\mu$m \citep{2009A&A...504..415S}, Herschel-SPIRE at 500, 350 and 250~$\mu$m \citep{2010A&A...518L...3G}, Herschel-PACS at 160 and 70~$\mu$m \citep{2010A&A...518L...2P}, MIPSGAL at 24~$\mu$m \citep{2009PASP..121...76C}, and the Spitzer-GLIMPSE at 8.0, 5.8, 4.5 and 3.6~$\mu$m \citep{2009PASP..121..213C}. The colorbar scale is linear, with units indicated in each subplot. }
\label{fig:background_IR_survey}
\end{figure*}

\begin{figure*}
        \centering
        \includegraphics[width=1.0\linewidth]{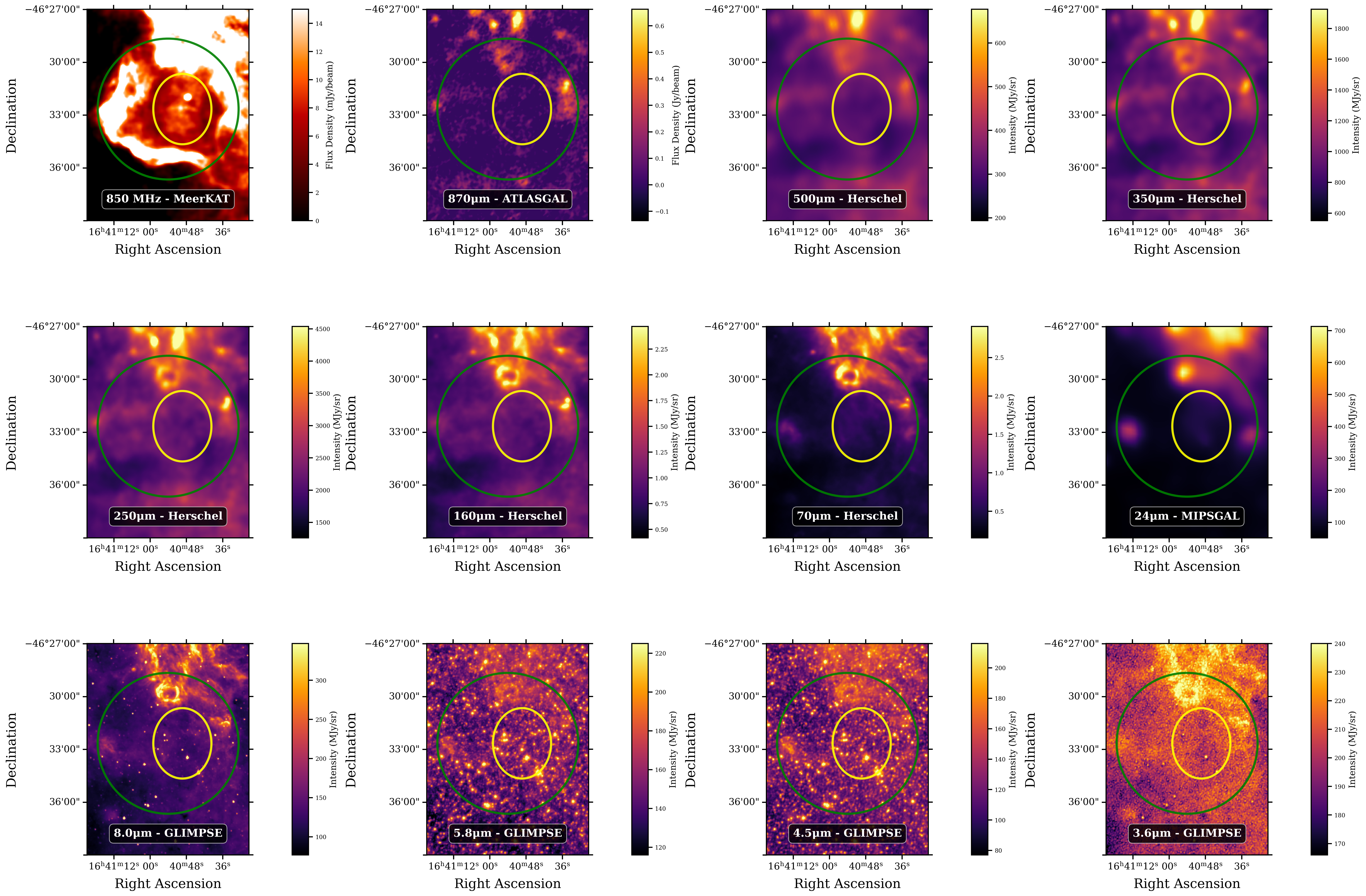}
\caption{ Multi-wavelength comparison of a zommed-in view of SNR G338.3$-$0.0 and the interior radio diffuse emission region.~The leftmost panel shows the MeerKAT 816 MHz image, with the interior radio diffuse emission (yellow ellipse), and SNR  G338.3$-$0.0 (green circle). The remaining panels display the corresponding FoV from MIR and FIR surveys, including ATLASGAL at 870~$\mu$m \citep{2009A&A...504..415S}, Herschel-SPIRE at 500, 350 and 250~$\mu$m \citep{2010A&A...518L...3G}, Herschel-PACS at 160 and 70~$\mu$m \citep{2010A&A...518L...2P}, MIPSGAL at 24~$\mu$m \citep{2009PASP..121...76C}, and the Spitzer-GLIMPSE at 8.0, 5.8, 4.5 and 3.6~$\mu$m \citep{2009PASP..121..213C}. The colorbar scale is linear, with units indicated in each subplot. }
\label{fig:pwn_IR_survey}
\end{figure*}

As discussed in \S\ref{sec:background}, we outline below the process for selecting a background region that satisfies the three criteria listed there.~Figures~\ref{fig:background_IR_survey} \& \ref{fig:pwn_IR_survey} present the MeerKAT 816~MHz radio continuum image alongside infrared maps spanning mid- to far-infrared (MIR to FIR) wavelengths. These include data from the APEX Telescope Large Area Survey of the Galaxy (ATLASGAL) at 870~$\mu$m \citep{2009A&A...504..415S}, the MIPSGAL survey at 24~$\mu$m \citep{2009PASP..121...76C}, and the Galactic Legacy Infrared Mid-Plane Survey Extraordinaire (GLIMPSE) at 3.6, 4.5, 5.8, and 8.0~$\mu$m \citep{2009PASP..121..213C}.~Additional IR data were obtained from the \textit{Herschel} space observatory using the Photodetector Array Camera and Spectrometer (PACS) at 70 and 160~$\mu$m \citep{2010A&A...518L...2P} and the Spectral and Photometric Imaging Receiver (SPIRE) at 250, 350, and 500~$\mu$m \citep{2010A&A...518L...3G}. We overlaid the regions corresponding to the interior radio diffuse emission (yellow ellipse) and SNR G338.3$-$0.0 (green circle) on each IR map.

As shown in Figure \ref{fig:pwn_IR_survey}, a low level of emission is detected from the interior of SNR G338.3$-$0.0 across the MIR and FIR bands.~This is consistent with previous analyses of this field, for example Figure 3 in the work by \cite{2011_castellati} and Figure 4 in the work by \cite{2016A&A...589A..51S}, which also show a lack of IR sources within the SNR, with thermal emission confined  to the northern shell region, as observed in Figures \ref{fig:thermal_sources} and \ref{fig:pwn_IR_survey}.
Therefore, to estimate the level of background emission inside the SNR, we choose a region of similarly low emission at these wavelengths, an example of which is shown in blue in Figure \ref{fig:background_IR_survey}.~To ensure the background region is sufficiently large that changes in the boundary do not result in significant variation in $\bar{\Sigma}_{\rm bkg}$, it encompasses areas with IR emission higher than the region of interest.  Such emission would likely increase $\bar{\Sigma}_{\rm bkg}$, resulting in a possible over-estimatation of the background emission within this SNR.

\subsection{Statistical Significance of the Interior Radio Diffuse Emission} 
\label{sec:appendix_stat_sign}

\begin{figure}[htbp]
    \centering
    \begin{subfigure}{0.45\textwidth}
        \centering
        \includegraphics[width=\linewidth]{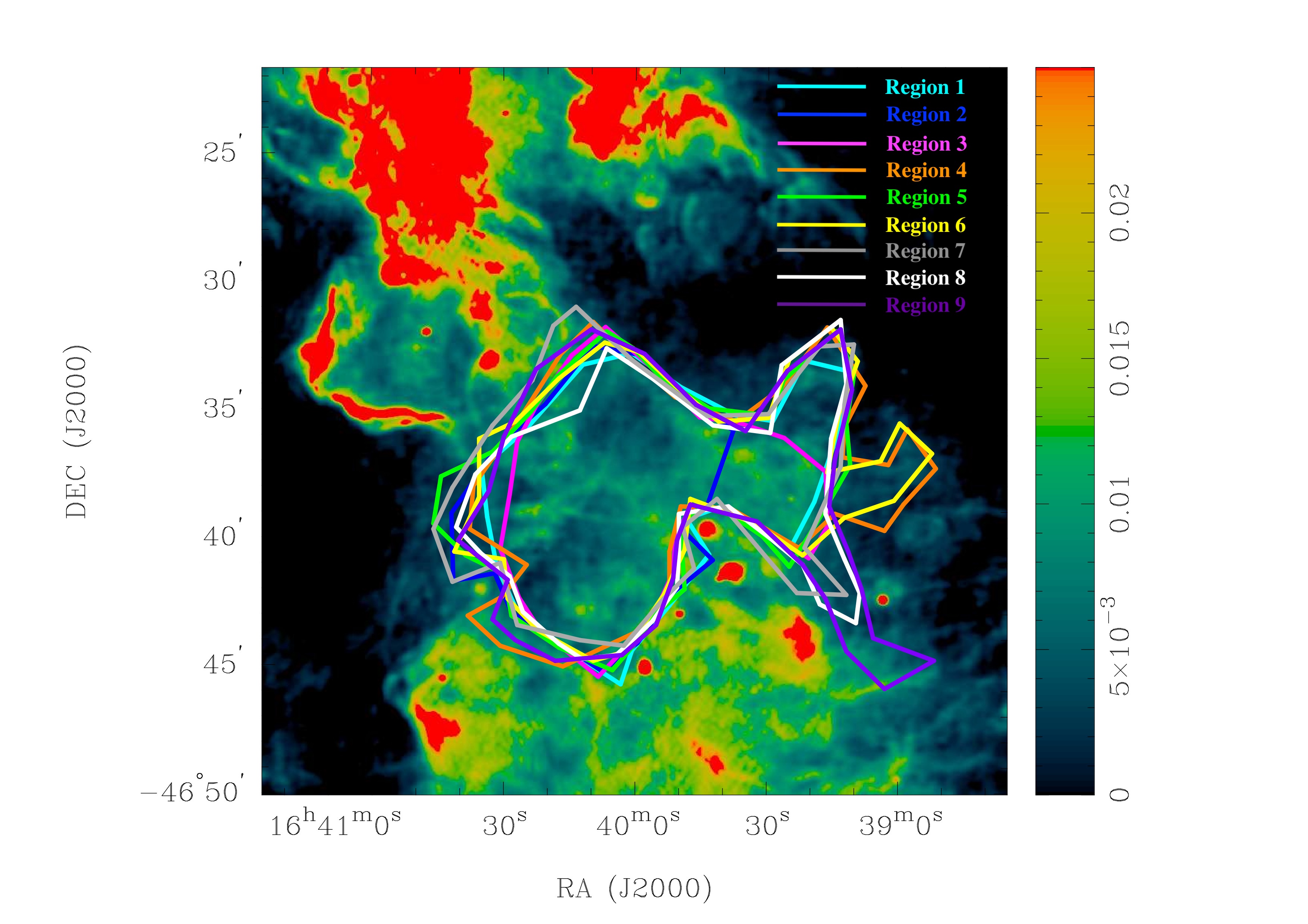}
        \caption{Location and extent of the 9 selected regions used to estimate the level  of the background emission in this field. Their corresponding flux density values are listed in Table \ref{tab:bkg_regions}.}
        \label{fig:background_regions}
    \end{subfigure}
    \hfill
    \begin{subfigure}{0.45\textwidth}
        \centering
        \includegraphics[width=\linewidth]{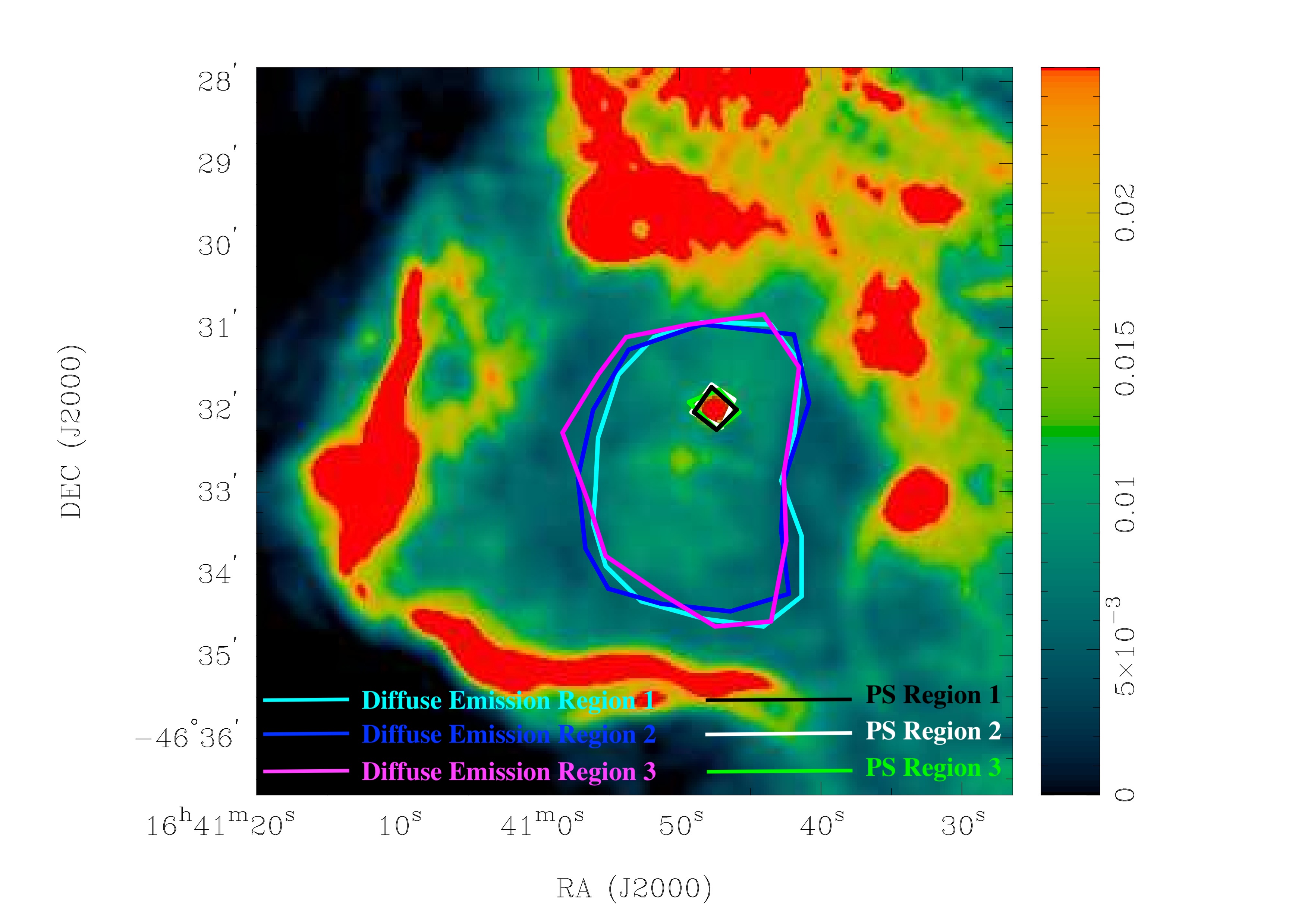}
        \caption{The different regions used to determine the flux density of the interior diffuse emission region and the PS. Their corresponding flux density values are listed in Table \ref{tab:raw_regions}.}
        \label{fig:pwn_regions}
    \end{subfigure}
    \caption{The selected (a) background, (b) PS and radio diffuse emission regions listed in Tables \ref{tab:bkg_regions} and \ref{tab:raw_regions}, respectively.~The intensity scale of the colormaps is linear and in the units of  Jy beam$^{-1}$.}
    \label{fig:all_regions}
\end{figure}

Here, we outline how we calculate each of the quantities listed in Equation \ref{eqn:delchi}.~To determine $\bar{\Sigma}_{\rm int, bkg}$, we first measured the total flux density $S_{\rm bkg}$ within the $\mathcal{N}_{\rm bkg}=9$ background regions shown in Figure~\ref{fig:background_regions}. The average surface brightness (Jy/pixel) within each background region $i$ is 
$\Sigma_{\rm bkg, i} \equiv \frac{S_{\rm bkg, i}}{N_{\rm pix,i}}$, where $S_{\rm bkg, i}$ is the total flux density within background region $i$ while $N_{\rm pix,i}$ is the number of image pixels it contains, as given in Table~\ref{tab:bkg_regions}. The average surface brightness of the overall background is then:

\begin{eqnarray}
    \bar{\Sigma}_{\rm bkg} & \equiv & \frac{1}{\mathcal{N}_{\rm bkg}} \sum\limits_{i=1}^{\mathcal{N}_{\rm bkg}} \Sigma_{\rm bkg, i},
\end{eqnarray}
with an uncertainty $\sigma_{\rm bkg}$:
\begin{eqnarray}
    \label{eq:rms_bkgg} 
    \sigma_{\rm bkg} & = &\sqrt{\frac{1}{\mathcal{N}_{\rm bkg}} \sum\limits_{i=1}^{\mathcal{N}_{\rm bkg}} \left(\Sigma_{\rm bkg, i} - \bar{\Sigma}_{\rm bkg} \right)^2}
\end{eqnarray} 
with the resulting values provided in Table \ref{tab:sigma}.

\begin{table*}[tb]
\centering
\caption{Flux Density  measurements of the 9 background regions shown in Figure \ref{fig:background_regions} }
\begin{tabular}{c|c|c|c|c|c|c|c|c|c}
\toprule
\toprule
Parameter & Region 1  & Region 2  & Region 3 & Region 4  & Region 5  & Region 6  & Region 7  & Region 8  & Region 9 \\
\midrule
 $S_{\rm bkg}$ (Jy)  & 15.19  & 12.06  & 13.76  & 16.26  & 16.44  & 16.04  & 17.21 & 14.85  & 17.57 \\
\( N_{\mathrm{bkg}} \) & 103216 & 85256 & 91537  & 118857 & 91537 & 117124 & 121010 & 106707 & 126633\\
 $\Sigma_{bkg}$ & 0.000147 & 0.000141 & 0.000150 & 0.000136  & 0.000139 & 0.000137 & 0.000142 &0.000139 & 0.000139 \\
 \bottomrule
\end{tabular}
\tablecomments{ $S_{bkg}$ is the total measured flux density in the background region, \( N_{\mathrm{bkg}} \) is the total number of pixels in the background region and  $\Sigma_{bkg}$ is the flux density per pixel in each bacgkround region.}
\label{tab:bkg_regions}
\end{table*}

\begin{table*}[htbp]
\caption{Flux Density  measurements for different enclosed regions of the SNR (Figure \ref{fig:image1}) , the SNR cut-out (Figure \ref{fig:image2}), the interior diffuse emission \& PS regions (Figure \ref{fig:pwn_regions}).}
\makebox[\textwidth][c]{
\setlength{\tabcolsep}{2pt}
\hspace*{-3cm}
\begin{tabular}{c|ccc|ccc|ccc|ccc}
\toprule
\toprule
Region & \multicolumn{3}{c}{SNR}  & \multicolumn{3}{c}{SNR cut-out}
  & \multicolumn{3}{c}{Interior Diffuse Emission}   & \multicolumn{3}{c}{PS}  \\
number & \( S_{\mathrm{SNR}}^{\mathrm{tot}} \)  (Jy)&  $N_{SNR}$  & $\Sigma_{SNR}$ & 
\( S_{\mathrm{SNR, cut}}^{\mathrm{tot}} \)  (Jy)   &$N_{SNR, cut}$  & $\Sigma_{SNR,cut}$  &
\( S_{\mathrm{int}}^{\mathrm{tot}} \)  (Jy) &$N_{int}$   & $\Sigma_{int}$&
  $S_{PS}$ (Jy) &$N_{PS}$ & $\Sigma_{PS}$  \\
\midrule  
1 & 13.26 & 45830 & 0.000289 & 11.11 & 30967 & 0.000359 & 1.34 & 7811  & 0.000172 & 0.0769 & 115 & 0.00067 \\
2 & 13.58 & 47093 & 0.000288 & 10.50 & 28834 & 0.000364 & 1.36  & 7935 & 0.000171 & 0.0809 & 127 & 0.00064 \\
3 & 13.56 & 46743 & 0.000290 & 11.18 & 30711 & 0.000364 & 1.37 & 7939  & 0.000173 & 0.0797 & 130 & 0.00061  \\
\bottomrule
\end{tabular}%
}
\tablecomments{$S_{SNR}^{tot}$ is the total measured flux density in each SNR region, \( N_{\mathrm{SNR}} \) is the total number of pixels within the SNR region, and $\Sigma_{SNR}$ is the flux density per pixel in each SNR region. $S_{SNR, cut}^{tot}$ is the total measured flux density in the SNR cut-out region, \( N_{\mathrm{SNR, cut}} \) is the total number of pixels in the SNR cut-out region and $\Sigma_{SNR, cut}$ is the flux density per pixel in each SNR cut-out region. \( S_{\mathrm{int}}^{\mathrm{tot}} \) is the total measured flux density in the interior diffuse emission region, \( N_{\mathrm{int}} \) is the total number of pixels in the interior diffuse emission region and $\Sigma_{int}$ is the flux density per pixel in each interior diffuse emission region. $S_{PS}$ is the total measured flux density in the PS region, \( N_{\mathrm{PS}} \) is the total number of pixels in the PS region and $\Sigma_{PS}$ is the flux density per pixel in each PS region.}
\label{tab:raw_regions}%
\end{table*}%

We followed a similar procedure to calculate the average surface brightness of the diffuse radio emission detected within this SNR $\bar{\Sigma}_{\rm int}$, where we calculated the total flux within several different regions.  However, as shown in Figure \ref{fig:pwn_regions}, this calculation needs to account for the contribution of the PS discussed in \S\ref{sec:UHF_band_analysis}.  To do so, we first measure the total flux density within $\mathcal{N}_{\rm int}=3$ regions corresponding to the diffuse interior emission (flux density $S_{{\rm int},i}^{\rm tot}$ within $N_{{\rm int},i}$ pixels ) and $\mathcal{N}_{\rm ps}=3$ regions around the PS (flux density $S_{{\rm PS},j}$ within $N_{{\rm PS},j}$ pixels), resulting in the values listed in Table \ref{tab:raw_regions}.  For each combination of interior and PS regions, we calculate the surface brightness of the interior diffuse emission:
\begin{eqnarray}
    \Sigma_{\rm int}^{i,j} & = & \frac{S_{{\rm int},i}^{\rm tot} - S_{{\rm ps},j}}{N_{{\rm int},i} - N_{{\rm PS},j}}.
\end{eqnarray}
The average surface brightness of the interior diffuse emission was calculated to be:
\begin{eqnarray}
    \bar{\Sigma}_{\rm int} & = & \frac{1}{\mathcal{N}_{\rm int}} \frac{1}{\mathcal{N}_{\rm PS}} \sum\limits_{i=1}^{\mathcal{N}_{\rm int}} \sum\limits_{j=1}^{\mathcal{N}_{\rm PS}} \Sigma_{\rm int}^{i,j},
\end{eqnarray}
with the uncertainty in this quantity defined as:
\begin{eqnarray}
    \sigma_{\rm int} & = & \sqrt{\frac{1}{\mathcal{N}_{\rm int}} \frac{1}{\mathcal{N}_{\rm PS}} \sum\limits_{i=1}^{\mathcal{N}_{\rm int}} \sum\limits_{j=1}^{\mathcal{N}_{\rm PS}}    \left(\Sigma_{\rm int}^{i,j} - \bar{\Sigma}_{\rm int} \right)^2},
\end{eqnarray}
resulting in the values given in Table \ref{tab:sigma}. Substituting these values into Equation \ref{eqn:delchi}, we find that $\Delta \chi \approx 5.22$.

\begin{figure*}[htbp]
    \centering
    \begin{subfigure}{0.45\textwidth}
        \centering
        \includegraphics[width=\linewidth]{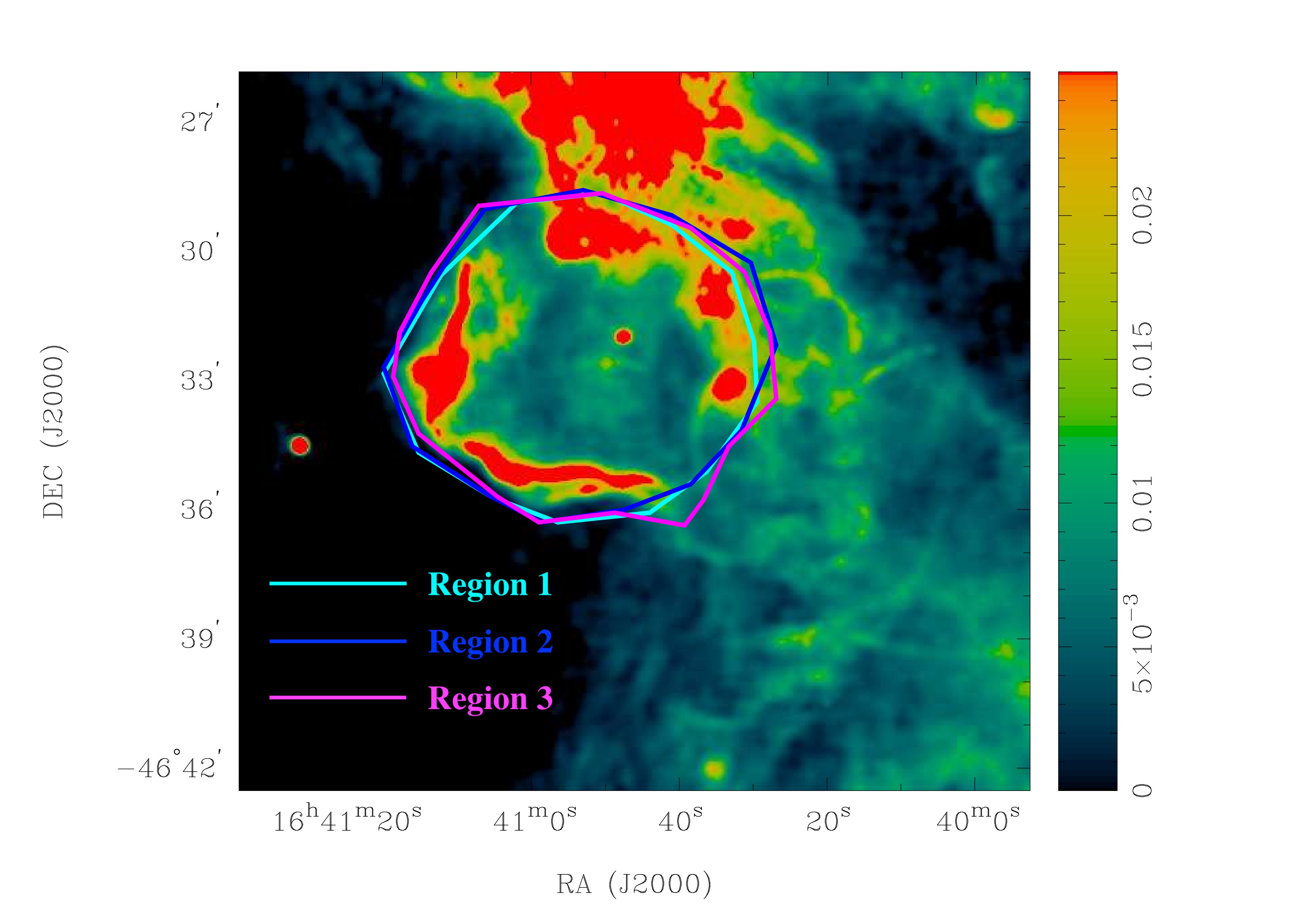}
        \caption{SNR Regions: Three slightly different regions used to enclose the SNR emitting region}
        \label{fig:image1}
    \end{subfigure}
    \hfill
    \begin{subfigure}{0.45\textwidth}
        \centering
        \includegraphics[width=\linewidth]{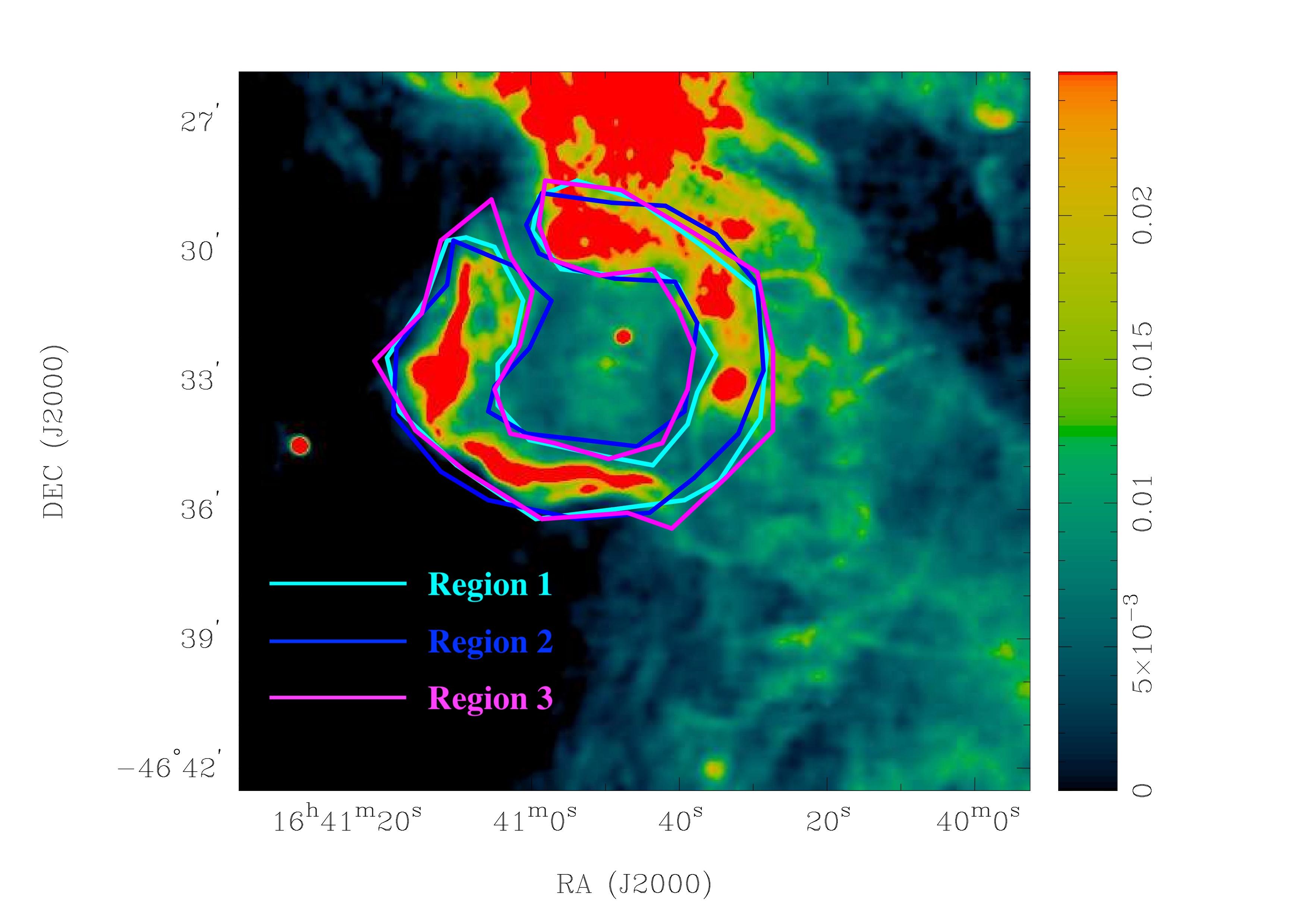}
        \caption{SNR cut-out Regions: Three slightly different regions used to enclose the SNR shell}
        \label{fig:image2}
    \end{subfigure}
    \caption{The selected (a) SNR regions and (b) SNR cut-out regions listed in Table~\ref{tab:raw_regions}. The intensity scale of the colormaps is linear and in units of Jy\,beam$^{-1}$.}
    \label{fig:all_regions}
\end{figure*}

\subsection{Flux Density Calculation For the Interior Radio Diffuse Emission Region} 
\label{sec:appendix_pwn_flux_density}

Following a similar procedure to that described for the surface brightness calculations in \S\ref{sec:source_sigma}, we subtract both the flux density of the PS ($S_{\mathrm{PS},j}$) and the estimated contribution from the background ($S_{\mathrm{bkg},k}$) scaled by the area of the interior regions, as follows:
\begin{align}
    S_{\mathrm{int}}^{i,j,k} &= S_{\mathrm{int, i}}^{\mathrm{tot}} - S_{\mathrm{PS, j}} - \left( S_{\mathrm{bkg, k}} \times \frac{N_{\mathrm{int, i}} - N_{\mathrm{PS, j}}}{N_{\mathrm{bkg, k}}} \right) \notag \\
    \label{eq:pwn_flux}
\end{align}

Where, $N_{\mathrm{int}, i}$, $N_{\mathrm{PS}, j}$, and $N_{\mathrm{bkg}, k}$ represent the number of pixels in the interior, point source, and background regions, respectively. All background regions are shown in Figure \ref{fig:background_regions}, while the PS and interior regions are highlighted in Figure \ref{fig:pwn_regions}.~Using the values in Tables \ref{tab:bkg_regions} and \ref{tab:raw_regions}, which correspond to the regions in Figures \ref{fig:background_regions} \& \ref{fig:pwn_regions} respectively, we obtain 81 estimates for the flux density of the interior diffuse emission, from which we compute an average value of \( \bar S_{\text{int}} = 180 \pm 34 \, \text{mJy} \).

\subsection{Flux Density Calculation for the SNR Shell} 
\label{sec:appendix_snr_flux_density}

To accurately determine the flux density of the SNR shell, we used two complementary methods, accounting for contributions from both the background and the interior diffuse emission region.~Each method offers a distinct approach to handling the contribution from the interior diffuse emission region. In Method I, we calculate the SNR shell flux density by subtracting the flux density of the interior radio diffuse emission region and the scaled background flux in the SNR emitting region, using:

\begin{align}
    S_{\mathrm{SNR}}^{i,j,k} &= S_{\mathrm{SNR, i}}^{\mathrm{tot}} - S_{\mathrm{int, j}}^{\mathrm{tot}} - \left( S_{\mathrm{bkg, k}} \times \frac{N_{\mathrm{SNR, i}} - N_{\mathrm{int, j}}}{N_{\mathrm{bkg, k}}} \right) \notag \\
    \label{eq:pwn_flux}
\end{align}

Where, $S_{\mathrm{SNR}, i}^{tot}$ is the total flux density in each SNR region shown in Figure \ref{fig:image1}, and listed in Table \ref{tab:raw_regions}. $S_{\mathrm{int}, j}^{tot}$ is the total flux density in each of the interior radio diffuse emission regions shown in Figure \ref{fig:pwn_regions} and listed in Table \ref{tab:raw_regions} and $S_{\mathrm{bkg}, k}$ is the total flux density in each of the background regions shown in Figure \ref{fig:background_regions} and listed in Table \ref{tab:bkg_regions}.~$N_{\mathrm{SNR}, i}$, $N_{\mathrm{int}, j}$, and $N_{\mathrm{bkg}, k}$ represent the number of pixels in the SNR, interior, and background regions, respectively.

In Method II, we measure the flux density in an ``SNR cut-out'' region that traces the SNR shell and excludes the PWN region (Figure \ref{fig:image2}), then subtract the scaled background flux for that specific region, using:

\begin{align}
    S_{\mathrm{SNR, cut}}^{i,j,k} &= S_{\mathrm{SNR, cut,  i}}^{\mathrm{tot}}  - \left( S_{\mathrm{bkg, j}} \times \frac{  N_{\mathrm{SNR, cut, i}}}{N_{\mathrm{bkg, j}}} \right) \notag \\
    \label{eq:pwn_flux}
\end{align}

Where, \( S_{\mathrm{SNR, cut, i}}^{\mathrm{tot}} \) is the total measured flux density within the SNR cut-out regions shown in Figure \ref{fig:image2} and listed in Table \ref{tab:raw_regions}.~$S_{\mathrm{bkg}, j}$ is the total flux density in each of the background regions shown in Figure \ref{fig:background_regions} and listed in Table \ref{tab:bkg_regions}.~$N_{\mathrm{SNR, cut}, i}$, and $N_{\mathrm{bkg}, j}$ represent the number of pixels in the SNR cut-out, and background regions, respectively. This method is particularly useful as it removes the need for a separate estimation of the flux density in the interior radio diffuse emission region. Using Method I, we computed 81 values, yielding an average SNR flux density of \( \bar S_{\text{SNR}} = 6.61 \pm 0.21 \, \text{Jy} \). Using Method II, we computed 27 values, yielding an average SNR flux density of \( \bar S_{\text{SNR, cut}} = 6.91 \pm 0.33 \, \text{Jy} \).~A weighted average derived from two distinct calculation methods yields a final SNR flux density of \( \bar S_{\text{SNR}} ^{Final} = 6.70 \pm 0.18 \, \text{Jy} \). 

\newpage

\bibliography{sample631}{}
\bibliographystyle{aasjournal}

\end{document}